\documentclass{article}
\usepackage[utf8]{inputenc}
\usepackage[left=3cm, right=3cm, top=2cm]{geometry}
\usepackage{xcolor}
\usepackage{hyperref}
\usepackage{tikz}
\usepackage{comment}
\usepackage{graphicx}
\usepackage{multirow}
\usepackage{colortbl}
\usepackage{authblk}

\definecolor{Gray}{gray}{0.9}

\usepackage{verbatim}
\usepackage[section]{placeins}
\usepackage{caption}
\usepackage{tabularx}
\usepackage[T1]{fontenc}
\usepackage{adjustbox}
\usepackage{diagbox}
\usepackage{booktabs}
\usepackage{multirow}
\usepackage{array}
\usepackage{makecell}
\usepackage[flushleft]{threeparttable}
\usepackage{amsmath}
\usepackage[section]{placeins}
\usepackage[super]{nth}
\usepackage{amssymb}
\usepackage{xcolor,colortbl}

\usepackage{fancyhdr}
\usepackage{smartdiagram}
\usepackage{metalogo}
\usepackage{setspace}
\usepackage[titletoc]{appendix}

\usetikzlibrary{calc,trees,positioning,arrows,chains,shapes.geometric,%
    decorations.pathreplacing,decorations.pathmorphing,shapes,shadows,%
    matrix,shapes.symbols}

\tikzset{
>=stealth',
  punktchain1/.style={
    rectangle, 
    rounded corners, 
    draw=black, very thick,
    minimum height=4em, 
    text centered, 
    on chain},
  line/.style={draw, thick, <-},
  small punktchain/.style={
  	rectangle, 
  	rounded corners, 
  	draw=black, very thick,
  	text width=5em, 
  	minimum height=3em, 
  	text centered, 
  	on chain},
  line/.style={draw, thick, <-},
  element/.style={
    tape,
    top color=white,
    bottom color=blue!50!black!60!,
    minimum width=8em,
    draw=blue!40!black!90, very thick,
    text width=10em, 
    minimum height=3.5em, 
    text centered, 
    on chain},
  every join/.style={->, thick,shorten >=1pt},
  decoration={brace},
  tuborg/.style={decorate},
  tubnode/.style={midway, right=2pt},
}
\usetikzlibrary{calc,
	trees,
	positioning,
	arrows,
	chains,
	shapes.geometric,%
	decorations.pathreplacing,
	decorations.pathmorphing,
	shapes,
	shadows,%
	matrix,
	shapes.symbols,
	shapes.multipart,
	arrows.meta,
	backgrounds
}

\tikzset{
>=stealth',
  punktchain/.style={
    rectangle, 
    rounded corners, 
    draw=black, very thick,
    minimum height=3em, 
    text centered, 
    on chain},
  line/.style={draw, thick, <-},
  small punktchain/.style={
  	rectangle, 
  	rounded corners, 
  	draw=black, very thick,
  	text width=5em, 
  	minimum height=3em, 
  	text centered, 
  	on chain},
  line/.style={draw, thick, <-},
  element/.style={
    tape,
    top color=white,
    bottom color=blue!50!black!60!,
    minimum width=8em,
    draw=blue!40!black!90, very thick,
    text width=10em, 
    minimum height=3.5em, 
    text centered, 
    on chain},
  every join/.style={->, thick,shorten >=1pt},
  decoration={brace},
  tuborg/.style={decorate},
  tubnode/.style={midway, right=2pt},
}
\usetikzlibrary{calc,
	trees,
	positioning,
	arrows,
	chains,
	shapes.geometric,%
	decorations.pathreplacing,
	decorations.pathmorphing,
	shapes,
	shadows,%
	matrix,
	shapes.symbols,
	shapes.multipart,
	arrows.meta,
	backgrounds
}

\tikzstyle{mypicture} = [bend angle=45,outer frame sep=1ex]

\tikzstyle{mylabel} = [inner sep=3pt,
fill=white,
line width = 0.1 pt,
label distance=10mm
,draw = black!50
]

\tikzstyle{labelnode}=[rectangle,
rounded corners=15,
draw=red!50,
fill=red!10,
font={\Huge},
inner sep=2pt,
minimum width=50mm,
minimum height=20mm,
align=center, 
thick]

\tikzstyle{mynode}=[ellipse,
draw=black!50,
font={\Large},
inner sep=2pt,
minimum width=60mm,
minimum height=40mm,
align=center, 
thick]

\tikzstyle{myarrow}=[
draw=black!50,
line width=5pt,
shorten >=5pt,
shorten <=5pt,
{}-{Latex[length=9mm]}]

\tikzstyle{labelnode}=[rectangle,
rounded corners=10,
outer color=orange!20,
inner color=white,
font={\Huge},
inner sep=2pt,
minimum width=66mm,
minimum height=18mm,
align=center, 
thick]

\tikzstyle{labelnodeR}=[labelnode,outer color=red!50,draw=black!50,fill=red!50,]
\tikzstyle{labelnodeG}=[labelnode,outer color=green!50,draw=black!50,fill=green!50]

\tikzstyle{mypoint}=[circle,
thick,
inner sep=0pt,
minimum size=8mm]

\tikzstyle{mypointR}=[mypoint,draw=black!50,fill=red!50]
\tikzstyle{mypointG}=[mypoint,draw=black!50,fill=green!50]

\tikzstyle{myflag}=[rectangle,
inner sep=2pt,
minimum width=90mm,
minimum height=10mm,
draw=black!60,
align=center, 
thick]

\tikzstyle{myflagempty}=[rectangle,
inner sep=2pt,
minimum width=20mm,
minimum height=10mm,
draw=none,
align=center, 
thick]

\tikzstyle{myline} = [thick,draw=black!75,fill=black!20,line width=6pt,]
\tikzstyle{transition} = [thick,draw=black!75,fill=black!20,line width=2pt,]
\providecommand{\keywords}[1]{\textbf{Index terms---} #1}

\pgfdeclarelayer{background}
\pgfsetlayers{background,main}
\title{\textbf{Bayesian Reliability Analysis 
Of The Power Law Process With Respect To The Higgins-Tsokos Loss Function For Modeling Software Failure Times}
}
\date{}
\author[1]{Freeh Alenezi}
\affil[1]{PhD Candidtae, Department of Mathematics and Statistics, \textit{ College of Arts and Sciences} \\
\textit{University of South Florida}\\ Tampa, 33647, USA}

\author[2]{Chris. Tsokos}
\affil[2]{Distinguished University Professor, Department of Mathematics and Statistics, \textit{ College of Arts and Sciences} \\
\textit{University of South Florida}\\ Tampa, 33647, USA}
    
\begin{document}

\maketitle

\doublespacing


\section{abstract}
The Power Law Process, also known as Non-Homogeneous Poisson Process, has been used in various aspects, one of which is the software reliability assessment. Specifically, by using its intensity function to compute the rate of change of a software reliability as time-varying function. Justification of Bayesian analysis applicability to the Power Law Process was shown using real data. The probability distribution that best characterizes the behavior of the key parameter of the intensity function was first identified, then the likelihood-based Bayesian reliability estimate of the Power Law Process under the Higgins-Tsokos loss function was obtained. As a result of a  simulation study and using real data, the Bayesian estimate shows an outstanding performance compared to the maximum likelihood estimate using different sample sizes. In addition, a sensitivity analysis was performed, resulting in the Bayesian estimate being sensitive to the prior selection; whether parametric or non-parametric.    

\keywords{Reliability growth; intensity function; non-homogeneous Poisson process; kernel density; loss function; robustness}

\section{Introduction}
\numberwithin{equation}{section}

Software reliability growth is often tested during the software development process to insure a good quality product. Repairable software is tested until a failure is detected, then fixed, and tested again until a new failure is detected. This reliability improvement of software has been studied for decades. Duane (1964) \cite{1} introduced the "learning curve approach", which is a plot of the failure rate (or the intensity function) of a system as a function of time. It is used to assess software reliability improvement over time. For example, software reliability has improved when we observe a negative curve, whereas a positive curve means that reliability is deteriorating. Stability in software reliability is achieved when there is no curve, i.e. the graph is a horizontal line. 
The number of failures in the interval $(0,t]$, $N(t)$, is considered a Poisson counting process after satisfying the following conditions \cite{44,412}:
\begin{enumerate}
\item $N(t=0)$=0.
\item Independent increment (counts of disjoint time intervals are independent). 
\item  It has an intensity function $V(t)$ = $\lim_{\Delta t\to 0}{\frac{P(N(t,t+\Delta t)=1)}{\Delta t}}$.
\item Simultaneous failures do not exist ($\lim_{\Delta t\to 0}. 
{\frac{P(N(t,t+\Delta t)=2)}{\Delta t}}$=0). 
\end{enumerate}
The probability of a random value $N(t)$=n is given by:

\begin{equation} \label{eq:2.1.1.PrFail1}
P(N(t) = n) = \frac {\exp \left \{ -\int_{0}^{t}V(t)dt  \right \} \left \{ \int_{0}^{t}V(t)dt \right \}^n }{n!}, \; t>0.
\end{equation}
\\
Crow (1974) proposed a non-homogeneous Poisson process (NHPP) \cite{3}, which is a Poisson process with a time-varying intensity function, given by:
 
\begin{equation} \label{eq:1.1.Poisson}
V(t)=V(t;\beta,\theta)=\frac{\beta}{\theta}\left(\frac{t}{\theta}\right)^{\beta-1}
,\; t>0, \; \beta>0, \; \theta>0,
\end{equation}\\
with $\beta$ and $\theta$ as the shape and scale parameters, respectively. This NHPP is also known as the power law process (PLP). 


The joint probability density function (PDF) of the ordered failure times $T_1$, $T_2$, ..., $T_n$ from an NHPP with intensity function $V(t;\beta,\theta)$ is given by:

\begin{equation} \label{m3}
f\left(t_{\mathrm{1}},..,t_n\right)\mathrm{=}\prod^n_{i\mathrm{=1}}{V\left(t_i;\beta,\theta\right)}exp\left\{\mathrm{-}\int^w_0{V\left(t;\beta,\theta\right)dt}\right\},
\end{equation}\\
where w is the so-called stopping time. Considering the failure truncation case (w = $t_n$), the conditional reliability function of the failure time $T_n$ given $T_1=t_1$, $T_2=t_2$, $T_3=t_3$,..., $T_{n-2} = t_{n-2}$, $T_{n-1} = t_{n-1}$ is a function of $V(t;\beta,\theta)$.

To monitor software reliability growth over time, an engineer can use the estimate of the $\beta$ value, the key parameter in the intensity function, since it plays a significant role during the testing process. For $\beta>1$, the number of failures would increase because the intensity function is increasing. On the other hand, if the intensity function is decreasing, $\beta < 1$ means that the number of failures would decrease, indicating improved software reliability. Note that in the case of a homogeneous Poisson process pertains when $\beta=1$,  in which case the intensity function will be $\frac{1}{\theta}$ and whatever changes have been made have had no effect on the outcome.

The NHPP has been used for analyzing software failure times, and for predicting the next failure event. Several publications show the effectiveness and usefulness of this model in assessing reliability growth \cite{2,444,4444,412}. In addition, NHPP has been used to study drug effectiveness in breast cancer treatment \cite{13} and in the formulation of a software cost model \cite{55}.

Since the intensity function is driving the NHPP, improving the existing methods to estimate the key parameter $\beta$ will certainly improve the accuracy of reliability growth assessment and help the structuring of maintenance strategies. Molinares and Tsokos \cite{41}, obtained a Bayesian estimate of the parameter $\beta$ and compared it with its approximate maximum likelihood estimate (MLE). The authors derived the Bayesian estimates with respect to squared-error loss function, using  Burr, Jeffreys, and inverted gamma probability distributions as the prior PDFs for $\beta$.

In performing Bayesian analysis on a real world problem, we need some sort of justification for pursuing this particular type of analysis. Once we have identified the probability distribution that characterized the probabilistic behavior of the failure time, we need to identify the prior PDF of $\beta$ and a loss function. The squared-error loss function is the most popular loss function used in Bayesian analysis because of its analytical tractability. 
It places a small weight on the estimates around the true value, but proportionally more weight on estimates far from the true value. Higgins and Tsokos \cite{441} proposed a new loss function that places exponential weight on extreme deviations from the true value, while remaining mathematically tractable.   

In the present study, we investigate the effectiveness of Bayesian analysis in using the Higgins-Tsokos (H-T) loss function (that puts the loss at the end of the process) for modeling software failure times. To accomplish this, we use the NHPP as the underlying failure distribution subject to using the Burr PDF as a prior of $\beta$. In addition, we utilize the H-T loss function to perform sensitive analysis of prior selections. We employ parametric and non-parametric priors, namely Burr, inverted gamma, Jeffery, and two kernel PDFs. Therefore, the primary objective of the study is to answer the following questions within a Bayesian framework:
 \begin{enumerate}
 \item What is the performance of the Bayesian estimate of $\beta$ under the H-T loss function compared to its MLE when modeling software failure times using PLP?

\item Is the Bayesian estimate of  $\beta$, using the H-T loss function in the PLP, sensitive to the selection of the prior PDF, both parametric and non-parametric?
\end{enumerate}
The paper is organized as follows: Section 2 describes the theory and development of the Bayesian reliability model; Section 3 presents the results and discussion; Section are is the conclusions.

\section{Theory and Bayesian Estimates}
\subsection{Review of the Analytical Power Law Process}
\numberwithin{equation}{subsection}
\renewcommand{\theequation}{\thesubsection.\arabic{equation}}

The probability of achieving $n$ failures of a given system in the time interval $(0, t]$ can be written as, \cite{41,412}:

\begin{equation} \label{eq:2.1.1.PrFail}
P(x = n; t) = \frac {\exp \left \{ -\int_{0}^{t}V(t;\beta,\theta)dt  \right \} \left \{ \int_{0}^{t}V(t;\beta,\theta)dt \right \}^n }{n!}, \; t>0,
\end{equation} 
where $V(t;\beta,\theta)$ is the intensity function given by (\ref{eq:1.1.Poisson}). The reduced expression

\begin{equation} \label{eq:2.1.2.RPrFail}
P(x = n; t) = \frac{1}{n!} \exp\left \{ -\frac{t}{\theta}^{\beta}\right \} \frac{t}{\theta}^{n\beta},
\end{equation} 
is the PLP that is commonly known as the Weibull or NHPP.

If the PLP is the underlying failure model of the failure times $t_1$, $t_2$, $t_3$,... , $t_{n-1}$, and $t_n$, the conditional reliability function of $t_n$ given $t_1$, $t_2$, $t_3$,... , $t_{n-1}$ can be written as, \cite{41,412}:

\begin{equation} \label{eq:2.1.3.CondRel}
R(t_n | t_1, t_2, ..., t_{n-1}) = \exp \left \{ \int_{t_{n-1}}^{t_{n}}  V(t;\beta,\theta)dt\right \}, \: tn > t_{n-1} > 0,
\end{equation} 
since it is independent of $t_1$, $t_2$, $t_3$, ... , $t_{n-2}$. \\ Since the reliability function, equation (\ref{eq:2.1.3.CondRel}), is written mathematically as a function of the intensity function, estimating the parameter $\beta$ in the $V(t;\beta,\theta)$ leads to estimation of the reliability function.  \\
The (MLE) of $\beta$ is a function of the largest failure time and the MLE of $\theta$ is also a function of the MLE of $\beta$. Let $T_1$, $T_2$, ..., $T_n$ denote the first $n$ failure times of the PLP, where $T_l < T_2 < ... < T_n$ are total times since the initial startup of the system. Thus, the truncated conditional PDF, $f_i(t | t_1, ... , t_{i-1})$, in the Weibull process and is given by,  \cite{41,412}:

\begin{equation} \label{eq:2.1.4.WeibullP}
f_i(t | t_1, ... , t_{i-1}) =\frac{\beta}{\theta}\left(\frac{t}{\theta}\right)^{\beta-1} \exp \left \{ -\frac{t}{\theta}^{\beta}+\frac{t_{i-1}}{\theta}^{\beta} \right \}, \: t_{i-1} < t.
\end{equation} 
With $t = (t_1, t_2, ... , t_n)$, the likelihood function for the first $n$ failure times of the PLP $T_1 = t_1, T_2 = t_2, ... , T_n = t_n$ can be written as: 

\begin{equation} \label{eq:2.1.5.Likelihood}
L(t;\beta)=\exp \left( - \left( \frac{t_n}{\theta}  \right) ^{\beta}  \right)
\left( \frac{\beta}{\theta}  \right)^n
\prod^n_{i=1} \left( \frac{t_i}{\theta} \right)^{\beta-1}.
\end{equation}\\

The MLE for the shape parameter is given by, \cite{2,3,41,412}:

\begin{equation} \label{eq:2.1.6.ShapeP}
\hat{\beta}_n=\frac{n}{\sum^{n}_{i=1}\log\left(\frac{t_n}{t_i}\right)},
\end{equation} 
and for the scale parameter is: 

\begin{equation} \label{eq:2.1.7.ScaleP}
\hat{\theta}_n=\frac{t_n}{n^{1/\hat{\beta}_n}}.
\end{equation}

Note that the MLE of $\theta$ depends on the MLE of $\beta$ using the largest (last) observed failure time. 

\subsection{Development of the Bayesian Estimates} 
\numberwithin{equation}{subsubsection}
\renewcommand{\theequation}{\thesubsection.\arabic{equation}}

Crow \cite{2,3} failure data from a system undergoing developmental testing was used, by Molinares \& Tsokos \cite{41}, to show how $\beta$ varied depending on the last failure time (largest time), thus they proposed a Bayesian approach to the PLP. The authors also found that the MLE of $\beta$ follows
a four-parameter Burr probability distribution, $g(\beta; \alpha, \gamma, \delta, \kappa)$, known as the four-parameter Burr type XII probability distribution, with a PDF given by:

\begin{equation} \label{eq:2.2.3.Burrtype}
g_{B}(\beta)=g(\beta; \alpha, \gamma, \delta, \kappa)=\begin{cases}
\frac{
	\alpha \kappa \left( 
					   \frac{\beta-\gamma}{\delta} 
				  \right)^{\alpha-1}
	}
	{
	\delta \left(
				1+\left(
						\frac{\beta-\gamma}{\delta}
				  \right)^{\alpha}
		   \right)^{\kappa+1}
	},
 & \gamma \le \beta < \infty \vspace{.25cm} \\ 
0, & otherwise\\
\end{cases},
\end{equation} \\
where the hyperparameters $\alpha$, $\gamma$, $\delta$ and $\kappa$ are being estimated using MLE in the goodness of fit (GOF) test applied to the $\beta$ estimates.
The Crow successive failure data for his system is given in Table \ref{times}. 
According to the reliability growth failure data, the system failed for the first time at $0.7$ units of time, $t_1 = 0.7$, and it failed the \nth{40} time at $3256.3$ units of time, $t_{40} = 3256.3$. The MLE of the parameter $\beta$ for $n=40$ is, \cite{41,412}:

\begin{equation} \label{eq:2.2.1.40MLE}
\hat{\beta}_{40}=\frac{40}{\sum^{40}_{i=1}\log\left(\frac{3256.3}{t_i}\right)} \simeq 0.49.
\end{equation}

If $\beta$ were treated in a non-Bayesian setting, its MLE would be given by Eq. (\ref{eq:2.2.1.40MLE}).

In an experimental process, the largest time to failure could occur at any point in the series of failures for a given system. Therefore, consider the case where the largest failure is $t_{39} = 3181$. In such a case, the estimate of $\beta_{39}$ is 0.48.


The largest failure time always affects the MLE of $\beta$. Thus, it is recommended that $\beta$ not to be thought of as an unknown constant \cite{41}, but rather as an unknown random variable. This recommendation provides the opportunity to study Bayesian analysis in the PLP with respect to various selections of loss functions and priors.

\begin{table}[t]
\centering
\captionsetup{  font=footnotesize,
  justification=raggedright,
}
\caption{Crow's failure times of a system under development.}
{\begin{tabular}{@{}lllllll@{}}
\toprule 
\multicolumn{7}{c}{Failure times}\\ \toprule
0.7\hphantom{000}   & 3.7\hphantom{000}   & 13.2\hphantom{000}  & 17.6\hphantom{000}  & 54.5\hphantom{000}  & 99.2\hphantom{000}  & 112.2\hphantom{00} \\
    120.9\hphantom{00} & 151\hphantom{000}   & 163\hphantom{0000}   & 174.5\hphantom{00} & 191.6\hphantom{00} & 282.8\hphantom{00} & 355.2\hphantom{00} \\
    486.3\hphantom{00} & 490.5\hphantom{00} & 513.3\hphantom{00} & 558.4\hphantom{00} & 678.1\hphantom{00} & 688\hphantom{000}   & 785.9\hphantom{00} \\
    887\hphantom{000}   & 1010.7\hphantom{0} & 1029.1\hphantom{0} & 1034.4\hphantom{0} & 1136.1\hphantom{0} & 1178.9\hphantom{0} & 1259.7\hphantom{0} \\
    1297.9\hphantom{0} & 1419.7\hphantom{0} & 1571.7\hphantom{0} & 1629.8\hphantom{0} & 1702.4\hphantom{0} & 1928.9\hphantom{0} & 2072.3\hphantom{0} \\
    2525.2\hphantom{0} & 2928.5\hphantom{0} & 3016.4\hphantom{0} & 3181\hphantom{00}  & 3256.3\hphantom{0} &  --\hphantom{0}     & --\hphantom{0} \\
\bottomrule
\end{tabular}} \label{times}
\end{table}

The Bayesian estimates of $\beta$ will be derived using H-T loss functions.

\subsubsection{Bayesian Estimates Using the Higgins-Tsokos Loss Function}
The H-T loss function (1976) is given by, \cite{41}:

\begin{equation} \label{eq:2.2.2.Higgins-Tsokos}
L(\hat{\xi},\xi) = \frac{f_1 \exp \left \{f_2 (\hat{\xi}-\xi) \right \}+f_2 \exp \left \{-f_1(\hat{\xi}-\xi) \right \} }{f_1 + f_2}-1, \: f_1, f_2 > 0.
\end{equation} 

Higgins and Tsokos \cite{441} showed that it places more weight on the extreme underestimation and overestimation of the true value when $f_1>f_2$ and $f_1<f_2$, respectively. The risk using the H-T loss function, where $\xi$ =$\beta$ represents the estimate of $\hat{\xi}$ =$\hat{\beta}$, is given by:

\begin{equation} \label{eq:2.2.2.Risk.Higgins-Tsokos}
E[L(\hat{\beta},\beta)] = \int^{\infty}_{-\infty} [\small{\frac{f_1 \exp \left \{f_2 (\hat{\beta}-\beta) \right \}+f_2 \exp \left \{-f_1(\hat{\beta}-\beta) \right \} }{f_1 + f_2}-1}] h(\beta|t) d\beta.
\end{equation} \\

By differentiating $E[L(\hat{\beta},\beta)]$ with respect to $\beta$ and setting  it equal to zero we solve for $\hat{\beta}$, the Bayesian estimate of $\beta$ with respect to the H-T loss function, is given by:

\begin{equation} \label{eq:2.2.2.3.HT}
\hat{\beta}_{B.TH}= \frac{1}{f_1+f_2} \ln[ \frac{\int^{\infty}_{-\infty} \exp \left \{f_1 \beta \right \}h(\beta|t) d\beta}{\int^{\infty}_{-\infty} \exp \left \{-f_2 \beta \right \}h(\beta|t) d\beta} ].
\end{equation} \\

The Bayesian estimate of $\beta$ with respect to the H-T loss function and Burr probability distribution, as the prior, has $h(\beta|t)$ given by:

\begin{equation} \label{eq:2.2.7.B.TH.Burr}
h(\beta|t) = \frac{
\int^{\infty}_{\gamma} (\frac{\beta}{\theta})^n 
\exp \left \{ -\left(\frac{t_n}{\theta} \right)^\beta \right \} 
\prod^n_{i=1} \left( \frac{t_i}{\theta} \right) ^{\beta-1} 
\frac{(\frac{\beta-\gamma}{\delta})^{\alpha-1}}
{(1+(\frac{\beta-\gamma}{\delta})^{\alpha})^{\kappa+1}} d\beta }{
\int^{\infty}_{\gamma} (\frac{\beta}{\theta})^n 
\exp \left \{ -\left(\frac{t_n}{\theta} \right)^\beta \right \} 
\prod^n_{i=1} \left( \frac{t_i}{\theta} \right) ^{\beta-1} 
\frac{(\frac{\beta-\gamma}{\delta})^{\alpha-1}}
{(1+(\frac{\beta-\gamma}{\delta})^{\alpha})^{\kappa+1}} d\beta}.
\end{equation} \\

With the use of Eq. (\ref{eq:2.1.3.CondRel}), the conditional reliability of $t_i$, the analytical structure of the conditional Bayesian reliability estimate for the PLP that is subject to the above information, is given by:

\begin{equation} \label{eq:BRelEstPLP}
\hat{R}_{B} (t_{i} | t_1, t_2, ..., t_{i-1}) = \exp
\left \{
-\int^{t_i}_{t_{i-1}} \hat{V}_{B}'(t;\beta,\theta)dt
\right \}, \; t_{i} > t_{i-1} > 0,
\end{equation}\\
where 
\begin{equation} \label{eq:HatVBt}
\hat{V}_{B}'(t;\beta_{B.TH},\theta) = \frac{\hat{\beta}_{B.TH}}{\theta}
\left( \frac{t}{\theta} \right)^{\hat{\beta}_{B.TH}-1}, \: \theta>0, t>0,
\end{equation} \\
where $\hat{\beta}_{B.TH}$ is the Bayesian estimate of $\beta$ using the H-T loss function. We are also interested in comparing the Bayesian estimate, using the H-T loss function, with MLE of the subject parameter for different parametric and non-parametric priors, assuming $\beta$ has a random behavior and $\theta$ is known; and also comparing Eq. (\ref{eq:2.1.7.ScaleP}) with an adjusted MLE considered as a function of $\beta$.

\subsection{Sensitivity Analysis: Prior Selection} \label{section:2.3}
\numberwithin{equation}{subsection}
\renewcommand{\theequation}{\thesubsection.\arabic{equation}}

In this section, we seek the answer to the following question: Is the Bayesian estimate of  $\beta$, using the H-T loss function in the PLP, sensitive to the selection of the prior, with parametric or non-parametric priors?
Assuming $\beta$ is a random variable, using simulated data, sensitivity analysis was done for the following parametric and non-parametric priors:

\begin{enumerate}
\item Jeffreys' prior [\cite{4443}]: \\
Jeffreys' prior is proportional to the square root of the determinant of the Fisher information matrix ($I(\beta$)). It is a non-informative prior, where the Jeffreys’ prior for the PLP, considering that $\beta$, $I(\beta)$ is scalar in this case, is given by:

\begin{equation} \label{eq:2.3.1.SqErLossF}
g_{J}(\beta) \propto\sqrt[]{I(\beta)}= \sqrt[]{-E(\frac{\partial^2 Log L(t;\beta)}{\partial \beta^2})}\propto \frac{1}{\beta}, \: \beta>0.
\end{equation} \\
\item The inverted gamma: \\
The PLP and inverted gamma probability distributions belong to the exponential family of probability distributions, which makes the latter a logical choice for an informative parametric prior for $\beta$. The inverted gamma probability distribution is given by:

\begin{equation} \label{eq:2.3.2.InvG}
g_{IG}(\beta) \propto \left( \frac{\mu}{\beta} \right)^{v+1} 
\frac{1}{\mu\Gamma(v)} \exp \left \{ \frac{-\mu}{\beta} \right \}, \: \beta>0, \mu>0, v>0,
\end{equation} 
where $v$ and $\mu$ are the shape and scale parameters.
\item Kernel' prior: \\
The kernel probability density estimation is a non-parametric method to approximately estimate the PDF of $\beta$ using a finite data set. It is given by:

\begin{equation} \label{eq:2.3.1.SqErLossF.ker}
g_{K}(\beta) = \frac{1}{nh} \sum_{i=1}^n K\bigg(\frac{\beta-\beta_i}{h}\bigg),
\end{equation} 
\end{enumerate} 
where $K$ is the kernel function and $h$ is a positive number called the bandwidth.
\subsubsection{The Jeffreys' Prior:}
\numberwithin{equation}{subsubsection}
\renewcommand{\theequation}{\thesubsubsection.\arabic{equation}}
Assuming Jeffreys' PDF, Eq. (\ref{eq:2.3.1.SqErLossF}), as the prior of $\beta$ and using the likelihood function (\ref{eq:2.1.5.Likelihood}), the posterior density of $\beta$ is given by:

\begin{equation} \label{eq:2.3.1.1.PostDens}
h_{J} (\bar{t} | \beta) = \frac
{\exp \left \{ \left(\frac{t_n}{\theta} \right)^{\beta} \right \}
\frac{\beta^{n-1}}{\theta^{n\beta}}
 \prod^n_{i=1} (t_i)^{\beta-1}}  
 {\int^{\infty}_0 
\exp \left \{ \left(\frac{t_n}{\theta} \right)^{\beta} \right \}
\frac{\beta^{n-1}}{\theta^{n\beta}}
\prod^n_{i=1} (t_i)^{\beta-1}d\beta}.
\end{equation}
\\ 
Thus, the Jeffreys' Bayesian estimate of $\beta$ in $V(t;\beta,\theta)$ under the 
H-T loss function, using Eq. (\ref{eq:2.2.2.3.HT}), is given by:

\begin{equation} \label{eq:2.3.1.2.JefBayEst.ht}
\hat{\beta}_{J.HT} = \frac{1}{f_1+f_2} \ln[ \frac{
\int^{\infty}_{\gamma}\exp \left \{f_1 \beta \right \} h_{J} (\bar{t} | \beta)  d\beta }{
\int^{\infty}_{\gamma} \exp \left \{-f_2 \beta \right \} h_{J} (\bar{t} | \beta) d\beta}].
\end{equation}
\\
We cannot obtain a closed analytical form of the Bayesian estimate, $\hat{\beta}_{J.HT}$, thus we must utilize numerical method to obtain the subject estimate. Also note that the estimate depends on knowing or being able to estimate the scale parameter $\theta$.

\subsubsection{The Inverted Gamma Prior:}
\numberwithin{equation}{subsubsection}
\renewcommand{\theequation}{\thesubsubsection.\arabic{equation}}

We proceed with our study with the prior probability density of $\beta$ given by the inverted gamma distribution Eq. (\ref{eq:2.3.2.InvG}). Using the likelihood Eq. (\ref{eq:2.1.5.Likelihood}), the posterior density of $\beta$ is given by:

\begin{equation} \label{eq:2.3.2.1.InvGamPr}
h_{IG}(t | \beta) = \frac
{\frac {\beta^{n-v-1}}{\theta^{n\beta}} 
\exp \left \{ -\left(\frac{t_n}{\theta}\right)^{\beta} - \frac{\mu}{\beta} \right \}
\prod^{n}_{i=1} (t_i)^{\beta-1}}
{\int_0^{\infty} \frac {\beta^{n-v-1}} {\theta^{n\beta}} 
\exp \left \{ -\left(\frac{t_n}{\theta}\right)^{\beta} - \frac{\mu}{\beta} \right \}
\prod^n_{i=1} (t_i)^{\beta-1} d\beta}.
\end{equation}
\\
Thus, the Bayesian estimate of $\beta$ under the inverted gamma distribution with respect to 
the H-T loss function, using Eq. (\ref{eq:2.2.2.3.HT}) and Eq. (\ref{eq:2.3.2.1.InvGamPr}), is given by:

\begin{equation} \label{eq:2.3.1.2.InvsBayEst.ht}
\hat{\beta}_{IG.HT}= \frac{1}{f_1+f_2} \ln[ \frac{
\int^{\infty}_{\gamma}\exp \left \{f_1 \beta \right \} h_{IG}(t | \beta) d \beta }{
\int^{\infty}_{\gamma} \exp \left \{-f_2 \beta \right \} h_{IG}(t | \beta) d \beta}].
\end{equation} \\
Here as well, we must rely on a numerical estimation of $\hat{\beta}_{IG.HT}$ because we cannot obtain a closed form of the above equation. Also note that the estimate depends on knowing or being able to estimate the scale parameter $\theta$.
\vspace{-3mm}
\subsubsection{The Kernel' Prior:}
\numberwithin{equation}{subsubsection}
\renewcommand{\theequation}{\thesubsubsection.\arabic{equation}}

Here, we shall assume the non-parametric kernel probability density Eq. (\ref{eq:2.3.1.SqErLossF.ker}) as the prior PDF of $\beta$; using the likelihood Eq. (\ref{eq:2.1.5.Likelihood}), the posterior density of $\beta$ is given by:

\begin{equation} \label{eq:2.3.1.1.PostDens.ker}
h_{k} (\bar{t} | \beta) = \frac
{\exp \left \{ \left(\frac{t_n}{\theta} \right)^{\beta} \right \}
\frac{\beta^{n}}{\theta^{n\beta}}
 \prod^n_{i=1} (t_i)^{\beta-1} \frac{1}{nh} \sum_{i=1}^n K\bigg(\frac{\beta-\beta_i}{h}\bigg)}  
 {\int^{\infty}_0 
\exp \left \{ \left(\frac{t_n}{\theta} \right)^{\beta} \right \}
\frac{\beta^{n}}{\theta^{n\beta}}
\prod^n_{i=1} (t_i)^{\beta-1}\frac{1}{nh} \sum_{i=1}^n K\bigg(\frac{\beta-\beta_i}{h}\bigg) d\beta}.
\end{equation} 

Thus, the kernel Bayesian estimate of the key parameter $\beta$ in $V(t;\beta,\theta)$ under the 
H-T loss function, using Eq. (\ref{eq:2.2.2.3.HT}) and Eq. (\ref{eq:2.3.1.1.PostDens}), is given by:

\begin{equation} \label{eq:2.3.1.2.JefBayEst.ht.ker}
\hat{\beta}_{K.HT} = \frac{1}{f_1+f_2} \ln[ \frac{
\int^{\infty}_{\gamma}\exp \left \{f_1 \beta \right \} h_{k} (\bar{t} | \beta) d\beta }{
\int^{\infty}_{\gamma} \exp \left \{-f_2 \beta \right \} h_{k} (\bar{t} | \beta) d\beta}].
\end{equation} 

We must rely on a numerical estimation because we cannot obtain a closed form solution for 
$\hat{\beta}_{K.HT}$. In addition, the kernel function, $K(u)$, and bandwidth, $h$, will be chosen to minimize the asymptotic mean integrated squared error (AMISE) given by:

\begin{equation} \label{amise.f1}
AMISE \left (\hat{f}(\beta) \right) = \int E \left [ \left (\hat{f}(\beta) - {f}(\beta) \right)^2 \right] d \beta,
\end{equation} \\
where $\hat{f}(\beta)$ and ${f}(\beta)$ are the estimated probability density of $\beta$ and the true probability density of $\beta$ respectively.
Below are details of the analysis we conducted using Monte Carlo simulation to generate data governed by a PLP, followed by using actual data.

\section{Results and Discussion}
\subsection{Numerical Simulation} \label{section:3.1}
\numberwithin{equation}{subsection}
\renewcommand{\theequation}{\thesubsection.\arabic{equation}}

A Monte Carlo simulation was used to compare the Bayesian (under H-T loss functions) and the MLE approaches. The parameter $\beta$ of the intensity function for the PLP was calculated using numerical integration techniques in conjunction with a Monte Carlo simulation to obtain its Bayesian estimate. Substituting these estimates in the intensity function, we obtained the Bayesian intensity function estimates, from which the reliability function can be estimated.

For a given value of the parameter $\theta$, a stochastic value for the parameter $\beta$ was generated from the Burr PDF. For each pair of values of $\theta$ and $\beta$, $400$ samples of $40$ failure times that followed a PLP were generated. This procedure was repeated $200$ times for three distinct values of $\theta$. The procedure is summarized in the algorithm (Algorithm \ref{alg1}) given below.

\begin{figure}[!htb]
\centering
\captionsetup{width=.9\textwidth,font=small   ,
  justification=raggedright}
   \resizebox{0.66\textwidth}{!}{\begin{minipage}{\textwidth}
\begin{tikzpicture}
  [node distance=.5cm,
  start chain=going below,]
     \node[punktchain,fill=orange!10, join,minimum width=2cm] (intro) {Start};
      \node[punktchain,fill=blue!3, join,minimum width=10cm] (init) {Initialize the parameter $\theta$ and number of iterations $p$};
     \node[punktchain, fill=blue!3,join,minimum width=10cm] (genb)      {Generate $\beta^{[k]}$ from Burr Distribution};
     \node[punktchain, fill=blue!3,join,minimum width=10cm] (gent)      {Generate $\vec{t}^{[k]}$ from PLP using $\beta^{[k]}$};
     \node[punktchain, fill=green!3,join,minimum width=10cm] (mleb) {Compute MLE of $\beta^{[k]}$, named $\hat{\beta}^{[k]}$};
     \begin{scope}[start branch=hoe]
      \node (iter) [punktchain,fill=purple!3, minimum width=1cm,on chain=going right] {\rotatebox{90}{$k=1,2,...,p$}};
     \end{scope}
     \node[punktchain, fill=green!3,join, minimum width=10cm] (baysb) {Compute Bayesian estimate, $\hat{\beta}^{[k]}_{B.HT}$, of $\beta^{[k]}$};
      \node [punktchain,fill=red!3, join,minimum width=10cm ] (mseb) {Calculate MSE of $\hat{\beta}_{B.HT}$};
  \node[punktchain, fill=red!3,join,minimum width=10cm] (msem) {Calculate MSE of $\hat{\beta}$ of $\beta$};
 \node[punktchain, ,fill=orange!8,join,minimum width=2cm] (makro) {End};
  \draw[-|,|-,-, thick,] (iter.south) |-+(0,-1.5em)-| (baysb.east);
   \draw[-|,|-,-, thick,] (iter.north) |-+(0,5.8em)-| (genb.east); 
  \end{tikzpicture}
 \vspace*{3pt}
 \captionsetup{labelformat=empty}
  	\caption{\textbf{Algorithm 1.} Simulation to analyze Bayesian estimates of $\beta$ for a given $\theta$.} \label{alg1}
    \end{minipage}}
\end{figure}
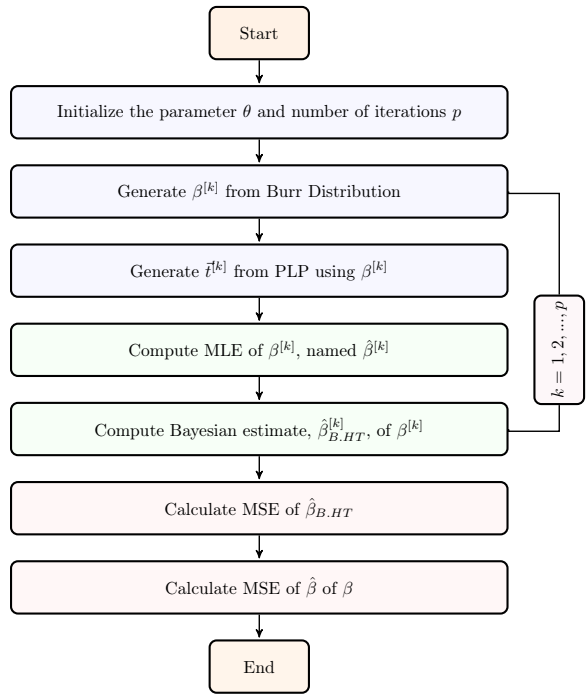

For each sample of size $40$, the Bayesian estimates and MLEs of the parameter were calculated when $\theta \in \{ 0.5, 1.7441, 4 \}$. The comparison is based on the mean squared error (MSE) averaged over the $100,000$ repetitions. The results are given in Table \ref{betaburr}.

\begin{table}[ht]
\centering
\captionsetup{width=.5\textwidth,font=footnotesize,
  justification=raggedright}
\caption{MSE for Bayesian estimates under the H-T loss function and MLE of $\beta$, for each assumed $\theta$ value.} 
{\begin{tabular*}{.5\linewidth}{@{\extracolsep{\fill}}lll} \toprule \label{betaburr}
\text{$\theta$} & \text{MSE of $\hat{\beta}$} & \text{MSE of $\hat{\beta}_{B.HT}$}  \\ \toprule
0.5\hphantom{0}   & 0.0112436\hphantom{0}    & 0.000507356\hphantom{0} \\
  1.7441\hphantom{0} & 0.0110573\hphantom{0}  & 0.000516057\hphantom{0} \\
     4\hphantom{0} & 0.010961\hphantom{0}   &   0.000518632\hphantom{0}\\
\bottomrule
\end{tabular*}}
\end{table}

It is observed that $\hat{\beta}_{B.HT}$ maintains a good accuracy, and is superior to $\hat{\beta}$ in estimating $\beta$ for the different values of $\theta$. For various sample sizes, the Bayesian estimate under the H-T loss function and the MLE of the parameter $\beta$ were calculated and averaged over $10,000$ repetitions. Table \ref{betaburravg} displays the simulated result of comparing a true value of $\beta$ with respect to its MLE and Bayesian estimates for $n=20$, $30$, ... , $160$. 

\begin{table}[ht]
\centering 
\captionsetup{width=.6\textwidth,font=footnotesize,
  justification=raggedright}
\caption{Bayesian estimates, under H-T loss function, and MLEs for the parameter $\beta$= 0.7054 averaged over 10,000
repetitions}
{\begin{tabular*}{.6\linewidth}{@{\extracolsep{\fill}}cccc} \toprule \label{betaburravg}
 \text{n} & \text{$\beta_{Fixed}$} & \text{$\hat{\beta}$} & \text{$\hat{\beta}_{B.HT}$} \\ \toprule
 20 & 0.7054 & 0.784026  & 0.675263 \\
 30 & 0.7054 & 0.756617  & 0.690189 \\
 40 & 0.7054 & 0.743982  & 0.696467 \\
 50 & 0.7054 & 0.73531\hphantom{0} & 0.699158 \\
 60 & 0.7054 & 0.729563  & 0.700642 \\
 70 & 0.7054 & 0.725977  & 0.70169\hphantom{0} \\
 80 & 0.7054 & 0.723338  & 0.702382 \\
 100 & 0.7054 & 0.719117  & 0.703165 \\
 120 & 0.7054 & 0.716315  & 0.703585 \\
 140 & 0.7054 & 0.714821  & 0.70398\hphantom{0} \\
 160 & 0.7054 & 0.713641  & 0.704244 \\
\bottomrule
\end{tabular*}}
\end{table}

\begin{figure}[ht]
\captionsetup{width=.6\textwidth,font=footnotesize,
  justification=raggedright}
	\centerline{\includegraphics[width=10cm] {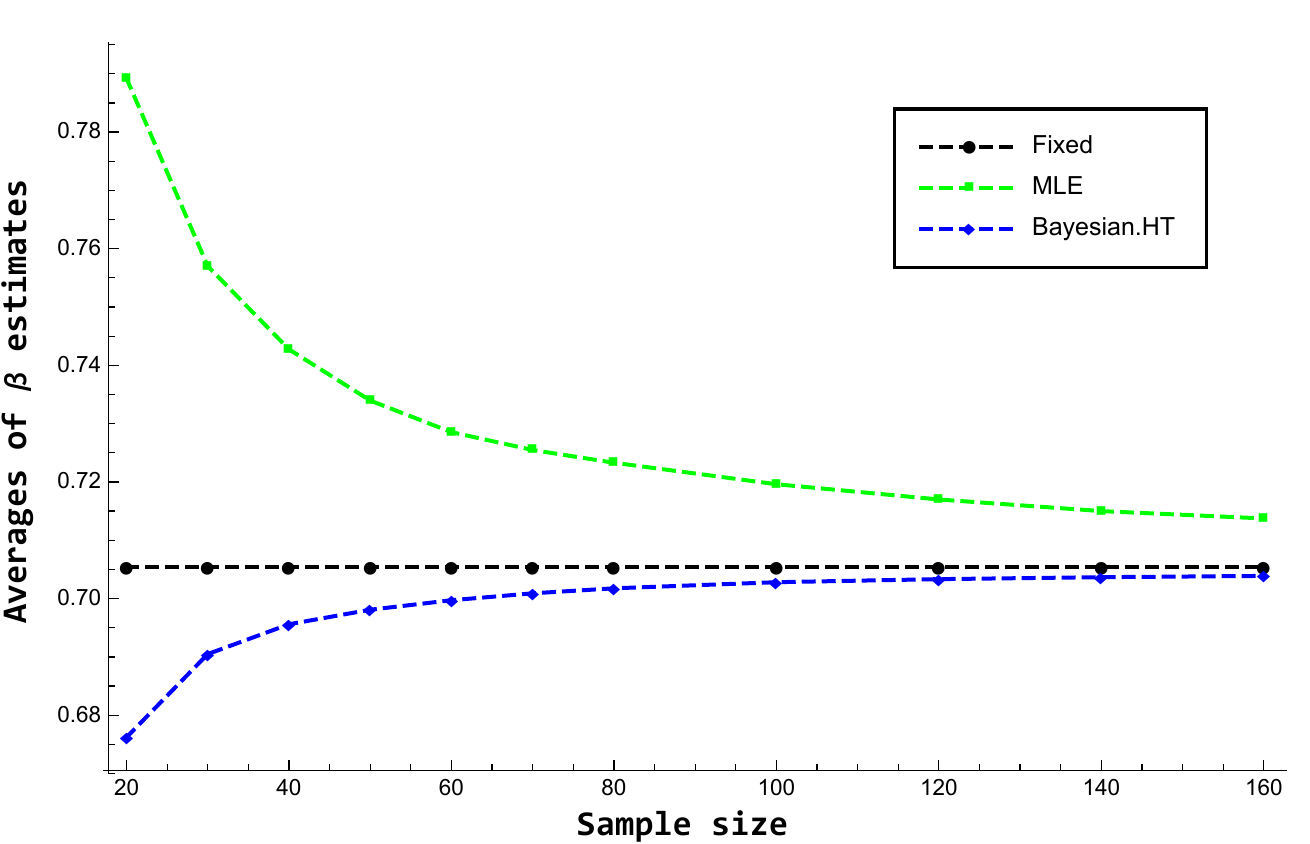}} 
	\vspace*{8pt}
	\caption{$\beta$ estimates versus sample size.}
    \label{betaburrss}
\end{figure}

Again, the Bayesian estimate is uniformly closer to the true value of $\beta$ than its MLE, even for a very small sample size of $n=20$. A graphical comparison of the true value of $\beta$ along with the Bayesian and MLE estimates as functions of sample size is given by Figure \ref{betaburrss}.

\begin{figure}[ht]
\captionsetup{width=.6\textwidth,font=footnotesize,
  justification=raggedright}
	\centerline{\includegraphics[width=10cm] {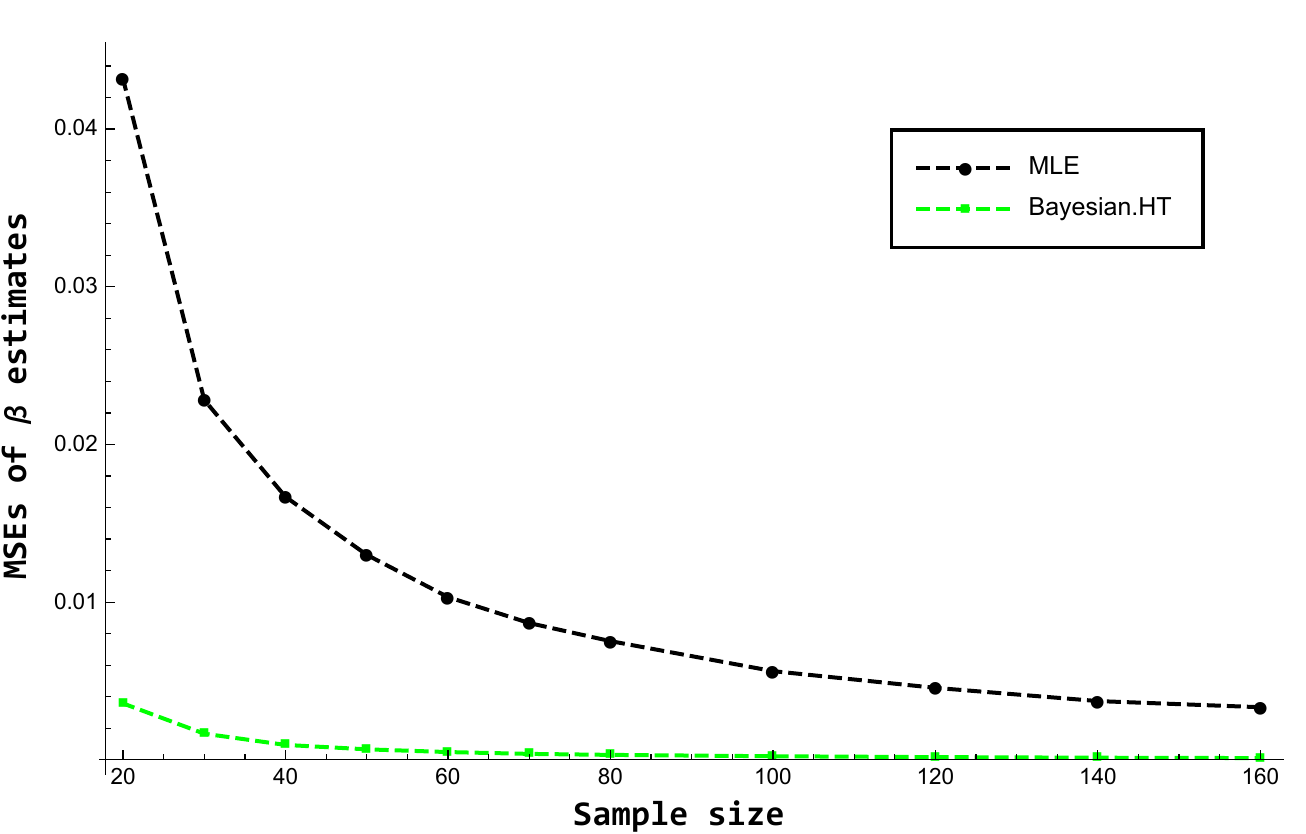}} 
	\vspace*{8pt}
	\caption{$\beta$ estimates versus sample size.}
    \label{msebetaburrss}
\end{figure}

Figure \ref{betaburrss} shows the MLE of $\beta$ tends to overestimate whereas the Bayesian estimate tends to underestimate the true value of $\beta$, particularly when considering small sample sizes. The MSEs of the MLE and Bayesian estimates of $\beta$ is given below by Figure \ref{msebetaburrss}. 

Regardless of sample size, the MSE of the Bayesian estimate of the key parameter $\beta$ is significantly smaller than the MSE of the MLE of $\beta$ ( Figure \ref{msebetaburrss}).

Since the Bayesian estimate under the H-T loss function for $\beta$ is better than its MLE, Molinares and Tsokos proposed to adjust the MLE of the parameter $\theta$ using (\ref{eq:2.1.6.ShapeP}) with a Bayesian estimate of $\beta$ instead of its MLE, both of which are needed to estimate the $V(t;\beta,\theta)$, as given below:

\begin{equation} \label{eq:3.1.1.MLEtetaAdj}
\hat{\theta}_{B.HT} = \frac{t_n}{n^{1/\hat{\beta_{B.HT}}}}
.\end{equation} 

For various sample sizes and the same $\beta$ ($\beta=0.7054$), the Bayesian MLE and MLE of the parameter $\theta$ and their corresponding MSEs were computed, averaging over the $10,000$ repetitions, using the MLE of $\theta$ ($\theta=1.7441$) of the Crow data.

\begin{table}[ht]
\centering
\captionsetup{width=.6\textwidth,font=footnotesize,
  justification=raggedright}
\caption{MLE and Bayesian estimates under the H-T loss function for the parameter $\theta$= 1.7441 averaged over 10,000
repetitions.} 
{\begin{tabular*}{.6\linewidth}{@{\extracolsep{\fill}}cccc}  \toprule \label{thetaburravg}
\text{n} & $\theta$  & \text{$\hat{\theta}_{MLE}$} & \text{$\hat{\theta}_{B.HT}$} \\ \toprule
 20 & 1.7441 & 3.17139  & 1.36422 \\
 30 & 1.7441 & 2.908\hphantom{00}  & 1.5097\hphantom{0} \\
 40 & 1.7441 & 2.73107  & 1.58115 \\
 50 & 1.7441 & 2.59245  & 1.61985 \\
 60 & 1.7441 & 2.48865  & 1.64406 \\
 70 & 1.7441 & 2.41782  & 1.66084 \\
 80 & 1.7441 & 2.36522 & 1.67294 \\
 100 & 1.7441 & 2.26774  & 1.68902 \\
 120 & 1.7441 & 2.20117  & 1.69923 \\
 140 & 1.7441 & 2.15539  & 1.70659 \\
 160 & 1.7441 & 2.11872  & 1.71193 \\
 \bottomrule
\end{tabular*}}
\end{table}

Table \ref{thetaburravg} shows the inferior performance for the MLE of $\theta$ and the slow convergence of its average values to $\theta=1.7441$, whereas the adjusted estimate of $\theta$ ($\hat{\theta}_{B.HT}$) using the Bayesian estimate of $\beta$ under the H-T loss function performed better in estimating the true value of $\theta$. The MLE of and the Bayesian MLE estimate of $\theta$ had a tendency to overestimate and underestimate the parameter $\theta$, respectively.

\begin{figure}[ht]
\captionsetup{width=.6\textwidth,font=footnotesize,
  justification=raggedright}
	\centerline{\includegraphics[width=10cm]{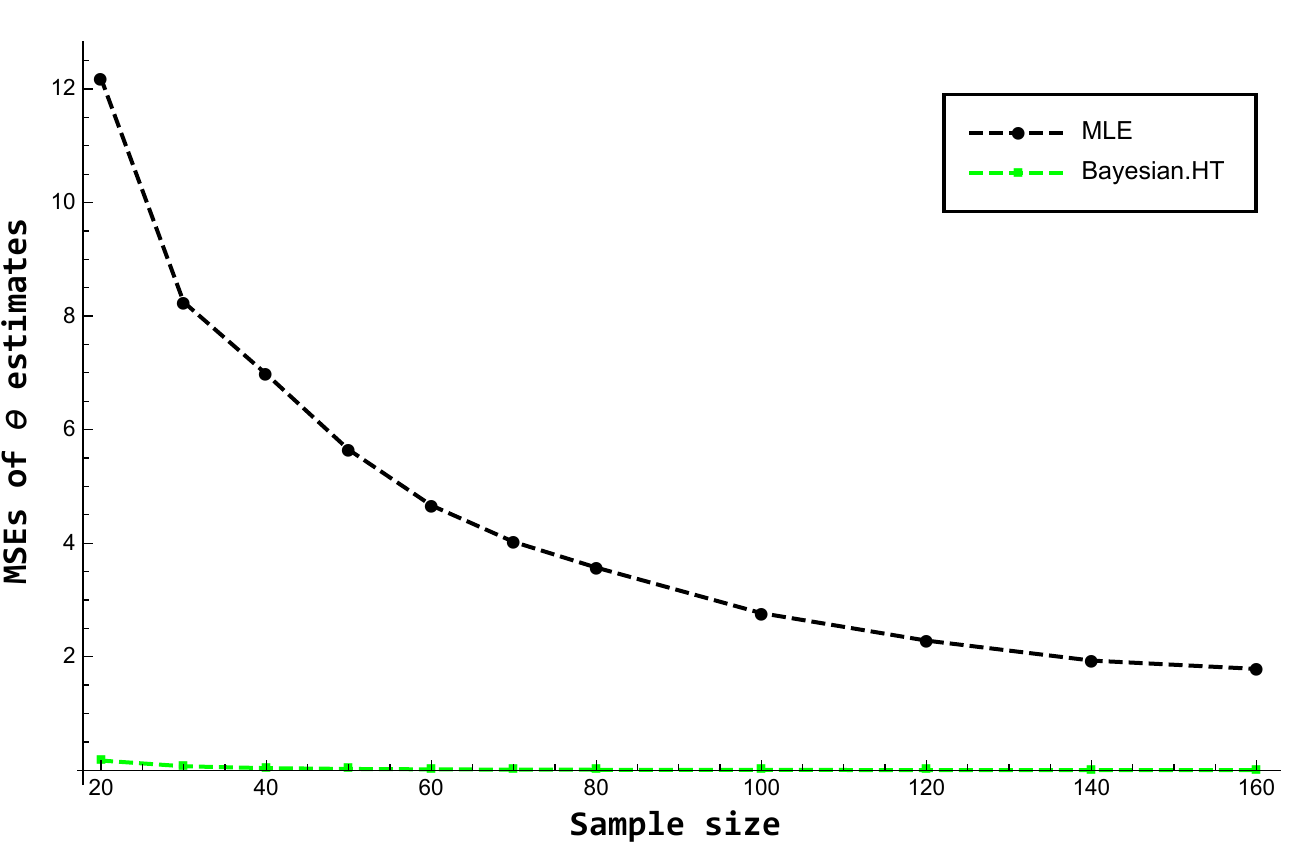}}
	\vspace*{8pt}
	\caption{MSE of $\theta$: Bayesian and MLE estimates versus sample size.}
    \label{thetaburrssmse}
\end{figure}

As expected, based on the Bayesian influence on $\beta$, $\hat{\theta}_{B.HT}$ is a better estimate than the MLE of $\theta$ ($\hat{\theta}$). This can be seen in Figure \ref{thetaburrssmse} where the MSEs of $\theta$ estimates were ploted against various sample sizes, which demonstrates the excellent performance of $\hat{\theta}_{B.HT}$.

We also computed the proposed estimate for the parameter $\theta$ ($\hat{\theta}_{B.HT}$) and its MSE over $100,000$ repetitions for different values of $\theta$  (0.5, 1.7441, 4) and sample size $n=40$. The results are given by Table \ref{3thetabetaburr}. The $\theta$ values (including 1.7441) were selected for this simulation are smaller and larger than the MLE of $\theta$ of the Crow data.

Table \ref{3thetabetaburr} below shows that the $\hat{\theta}_{B.HT}$ performed well for the selected $\theta$ values. This is particularly true for the small and medium value of $\theta$ values.

\begin{table}[ht]
\centering
\captionsetup{width=.6\textwidth,font=footnotesize,
  justification=raggedright}
\caption{MSE of $\theta$ estimates: Bayesian under the H-T loss function, and MLE of $\beta$.} 
{\begin{tabular*}{.6\linewidth}{@{\extracolsep{\fill}}ccc}  \toprule \label{3thetabetaburr}
$\theta$ & $\hat{\theta}_{B.HT}$  & MSE of $\hat{\theta}_{B.HT}$  \\ \toprule
0.5\hphantom{000}     & 0.503314\hphantom{0} & \textbf{0.00691164} \\
  1.7441  & 1.7509\hphantom{000} & \textbf{0.0827802\hphantom{0}} \\
     4\hphantom{0000}   &4.01025\hphantom{00} &  \textbf{0.439035\hphantom{00}}\\
\bottomrule
\end{tabular*}}
\end{table}

\begin{figure}
\captionsetup{width=.6\textwidth,font=footnotesize,
  justification=raggedright}	\centerline{\includegraphics[width=10cm]{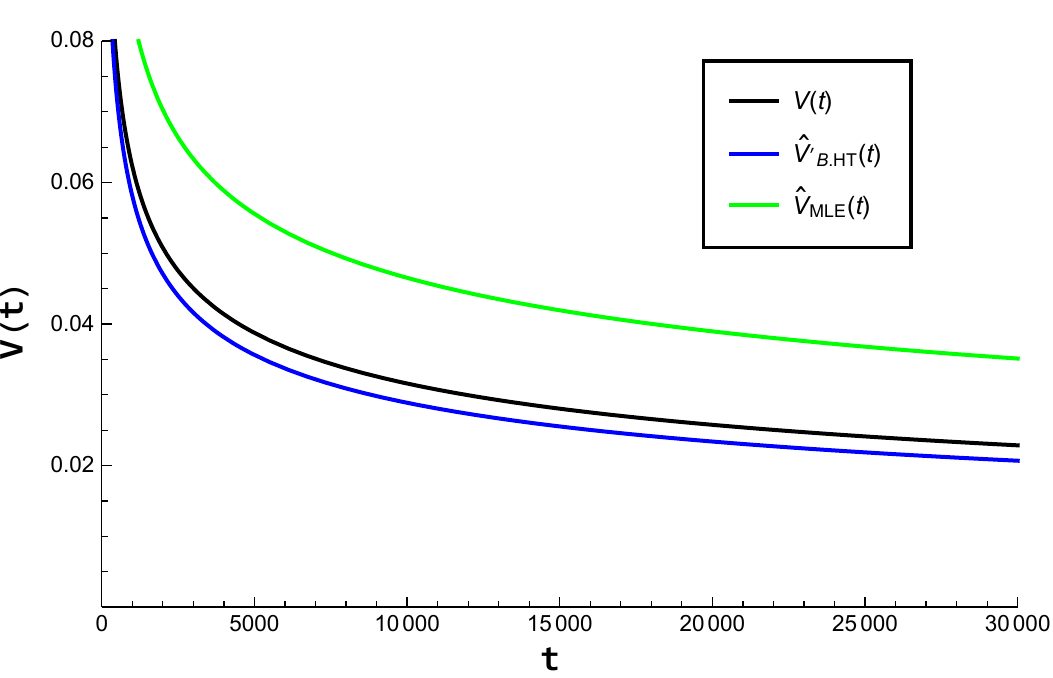}}
	\vspace*{8pt}
	\caption{Graph for $\theta$ = 1.7441 and the corresponding $\beta$ Bayesian estimate and MLE’s used in $\hat{V'}_{MLE}$ and $\hat{V'}_{B.HT}$, estimates of $V(t;\beta,\theta)$ with n = 40.}
    \label{fixedthetav}
\end{figure}

For a fixed value of $\theta=1.7441$ and a sample size similar to the size of the collected data, $n = 40$, the estimates of the intensity function $\hat{V}_{MLE}(t)$ and $\hat{V}_{B.HT}(t)$  were obtained using $\hat{\beta}$ and $\hat{\beta}_{B.HT}$, respectively, in Eq. (\ref{eq:1.1.Poisson}). That is,

\begin{equation}
\hat{V'}_{MLE}(t) = \frac{\hat{\beta}}{\theta}
\left( \frac{t}{\theta} \right)^{\hat{\beta}-1},\: \theta>0,  t>0.
\end{equation}
and
\begin{equation}
\hat{V'}_{B.HT} (t) = \frac{\hat{\beta}_{B.HT}}{\theta}
\left( \frac{t}{\theta} \right)^{\hat{\beta}_{B.HT}-1},\: \theta>0,  t>0.
\end{equation} 
Their graphs (Figure \ref{fixedthetav}) reveal the superior performance of $\hat{V'}_{B.HT}(t)$. \\

In order to obtain Bayesian estimates of the intensity function, $\hat{V^*}_{B.SE}$ and $\hat{V^*}_{B.HT}$, we substituted the Bayesian estimates of $\beta$ and its corresponding $\theta$ MLE in Eq. (\ref{eq:1.1.Poisson}). That is,

\begin{equation}
\hat{V^*}_{B.HT}(t) = \frac{\hat{\beta}_{B.HT}}{\hat{\theta}}
\left(\frac{t}{\hat{\theta}}\right)^{\hat{\beta}_{B.HT}-1},\: t>0.
\end{equation} 

The MLE of the intensity function, $\hat{V}_{MLE}$, is obtained using the MLEs of $\beta$ and $\theta$. That is,

\begin{equation}
\hat{V}_{MLE}(t) = \frac{\hat{\beta}}{\hat{\theta}}
\left(\frac{t}{\hat{\theta}}\right)^{\hat{\beta}-1},\: t>0.
\end{equation} 

The Bayesian MLE of the intensity function under the influence of the Bayesian estimate of $\beta$, denoted by $\hat{V}_{B.HT}$, is obtained by substituting $\hat{\beta}_{B.HT}$ and $\hat{\theta}_{B.HT}$ in Eq. (\ref{eq:1.1.Poisson}):

\begin{equation}
\hat{V}_{B.HT}(t) = \frac{\hat{ \beta}_{B.HT}}{\hat{\theta}_{B.HT}}
\left( \frac{t}{\hat{\theta}_{B.HT}} \right)^{\hat{\beta}_{B.HT}-1}, \: t>0.
\end{equation} \\

To measure the robustness of $\hat{V}_{B.HT}$ with respect to $\hat{V}_{MLE}$ , we calculated the relative efficiency (RE) of the estimate $\hat{V}_{B.HT}$ compared to the estimate $\hat{V}_{MLE}$, which is defined as

\begin{equation} \label{eq:3.1.2.ReletiveEff}
RE( \hat{V}_{B.HT}, \hat{V}_{MLE} ) = \frac{IMSE(\hat{V}_{B.HT})}{IMSE(\hat{V}_{MLE})} = \frac
{\int^{\infty}_{-\infty} [ \hat{V}_{B.HT} (t) - V (t)]^2dt}
{\int^{\infty}_{-\infty} [ \hat{V}_{MLE} (t) - V (t)]^2dt} \hspace{.1mm}.
\end{equation}

If $RE=1$, $\hat{V}_{B.HT}$ and $\hat{V}_{MLE}$ will be interpreted as equally effective. If $RE<1$, $\hat{V}_{B.HT}$ is more efficient than $\hat{V}_{MLE}$, contrary to $RE>1$, in which case $\hat{V}_{B.HT}$ is less efficient than $\hat{V}_{MLE}$.



Bayesian estimates and MLEs for the parameters  $\beta$= 0.7054 and $\theta$=1.7441 (Table \ref{values4v}), averaged over 10,000 repetitions, were used, for $n = 40$, to compare $\hat{V}_{B.TH}$ and $\hat{V}_{MLE}$ using Eq. (\ref{eq:3.1.2.ReletiveEff}).

\begin{table}[ht]
\centering
\captionsetup{width=.65\textwidth,font=footnotesize,
  justification=raggedright}\caption{Averages of the Bayesian (under the H-T loss function) and MLE estimates of $\beta$ and $\theta$} 
{\begin{tabular}{@{}ccccccc@{}} \toprule \label{values4v}
 $\beta$ & $\hat{\beta}$ & $\hat{\beta}_{B.HT}$ & $\theta$ & $\hat{\theta}$ & $\hat{\theta}_{B.HT}$  \\ \toprule
0.7054 & 0.743982 &  0.696467 & 1.7441 & 2.73107  & 1.58115 \\
\bottomrule
\end{tabular}}
\end{table}

Table \ref{values4v} above reveals that the averages of the Bayesian and Bayesian MLE estimates of the parameters $\beta$ and $\theta$ under the H-T loss function, respectively, are closer to the true values than their corresponding MLE estimates. The results of the comparison among the $\hat{V}_{B.TH}$ and $\hat{V}_{MLE}$ using Eq. (\ref{eq:3.1.2.ReletiveEff}) are given in Tables \ref{relativeV} and \ref{relativeVules}.

\begin{table}
\centering
\captionsetup{width=1\textwidth,font=footnotesize,
  justification=raggedright}\caption{Intensity functions with Bayesian and MLE estimates for $\beta$ and $\theta$ } 
{\begin{tabular}{@{}cccc@{}} \toprule \label{relativeV}
 $V(t)$ & $\hat{V}_{MLE}$ & $\hat{V^*}_{B.HT}$ & $\hat{V}_{B.HT}$ \\  \toprule
 $0.476465\cdot t^{-0.2946}$ & $0.352321\cdot t^{-0.256018}$ & $0.345946 \cdot t^{-0.303533}$ & $0.5062 \cdot t^{-0.303533}$  \\
\bottomrule
\end{tabular}}
\end{table}

The analytical forms of the $V(t)$, $\hat{V}_{MLE}$, $\hat{V^*}_{B.TH}$, and $\hat{V}_{B.TH}$ (Table \ref{relativeV}) were derived by substituting the initialized values, MLE estimates of both parameters, Bayesian estimate $\beta$ and MLE of $\theta$, and Bayesian estimates, respectively. 

\begin{table}[ht]
\centering
\captionsetup{width=.5\textwidth,font=footnotesize,
  justification=raggedright}
  \caption{Relative efficiency of $\hat{V}_{B.HT}$ compared to $\hat{V}_{MLE}$. } 
{\begin{tabular}{@{}ccc@{}} \toprule \label{relativeVules}
  $RE(\hat{V}_{B.HT},\hat{V}_{MLE})$ & $RE(\hat{V}_{B.HT},\hat{V^*}_{B.HT})$ \\ \toprule
  0.0761919\hphantom{0} & 0.00550275 \hphantom{0}\\
\bottomrule
\end{tabular}}
\end{table}

\begin{figure}[ht]
\captionsetup{width=.6\textwidth,font=footnotesize,
  justification=raggedright}
	\centerline{\includegraphics[width=10cm]{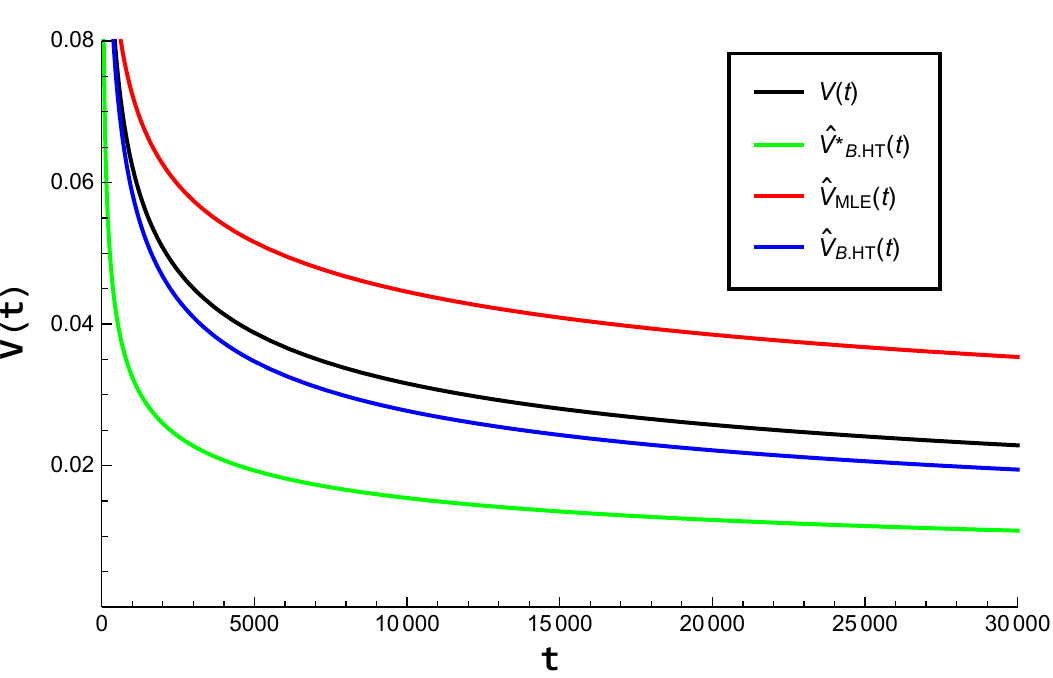}}
	\vspace*{8pt}
	\caption{Estimates of the intensity function using values in Table \ref{values4v}, n = 40.}
    \label{estimatedvalues}
\end{figure}

Table \ref{relativeVules} shows the comparison result of $\hat{V}_{B.HT}$ and $\hat{V}_{MLE}$, where the  $RE(\hat{V}_{B.HT},\hat{V}_{MLE})$ is less than 1, which implies that the intensity function using $\hat{\beta}_{B.HT}$ is more efficient than the intensity function under $\hat{\beta}_{MLE}$, establishing the superior relative efficiency of Bayesian estimates under the H-T loss function over MLE estimates. The corresponding graph for the intensity functions is given by Figure \ref{estimatedvalues}. In addition, $\hat{V^*}_{B.HT}$  computed using a Bayesian estimate for $\beta$ and MLE estimate for $\theta$, is less efficient compared to $\hat{V}_{MLE}$, and $\hat{V}_{B.HT}$.

The $\hat{V}_{B.HT}$ is a better estimate of $V(t)$, compared to the $\hat{V}_{MSE}$ and $\hat{V^*}_{B.HT}$. Based on the results of this section, the Bayesian estimates under the H-T loss function will be used to analyze real data in the following section.

\subsection{Using Real Data}

Using the software reliability growth data from Table \ref{times}, we computed $\hat{\beta}_{B.HT}$ and the adjusted estimate of $\theta$ ($\hat{\theta}_{B.HT}$) in order to obtain a Bayesian estimate of the intensity function under the H-T loss function. We followed the algorithm given below (Algorithm 2) to obtain the Bayesian intensity function for the given real data.

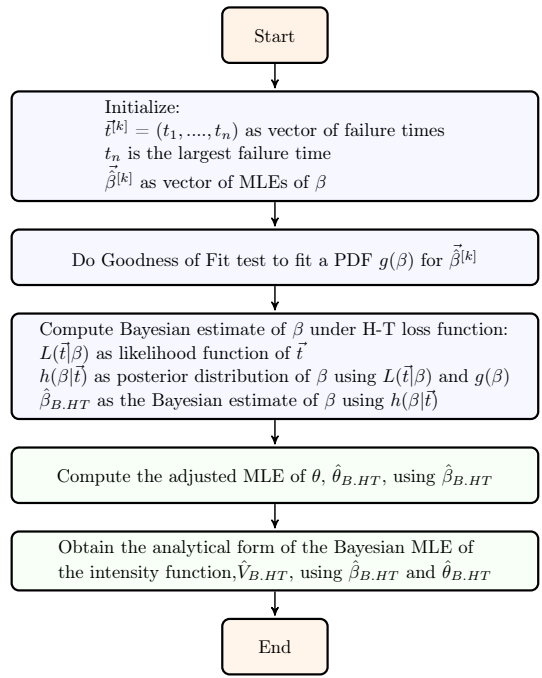
\begin{figure}[ht]
\centering
\begin{center}
\captionsetup{width=.9\textwidth,font=footnotesize,
  justification=raggedright}
   \resizebox{0.7\textwidth}{!}{\begin{minipage}{\textwidth}
\begin{tikzpicture}
  [node distance=.5cm,
  start chain=going below,]
     \node[punktchain,fill=orange!10, join,minimum width=2cm] (intro) {Start};
      \node[punktchain,fill=blue!3, join,minimum width=10cm] (init) {\begin{tabular}{l} Initialize: \\ $\vec{{t}}^{[k]}$ = $(t_1,....,t_n)$ as vector of failure times \\ $t_n$ is the largest failure time  \\ $\vec{\hat{\beta}}^{[k]}$ as vector of MLEs of $\beta$ \end{tabular}};
     \node[punktchain, fill=blue!3,join,minimum width=10cm] (genb)      {Do Goodness of Fit test to fit a PDF $g(\beta)$ for $\vec{\hat{\beta}}^{[k]}$ };
     \node[punktchain, fill=blue!3,join,minimum width=10cm] (gent)      {\begin{tabular}{l} Compute Bayesian estimate of $\beta$ under H-T loss function: \\ $L(\vec{t}| \beta)$ as likelihood function of $\vec{t}$ \\ $h(\beta | \vec{t})$ as posterior distribution of $\beta$ using $L(\vec{t}| \beta)$ and $g(\beta)$ \\ $\hat{\beta}_{B.HT}$ as the Bayesian estimate of $\beta$ using $h(\beta | \vec{t})$  \end{tabular}};
     \node[punktchain, fill=green!3,join,minimum width=10cm] (mleb) {Compute the adjusted MLE of $\theta$, $\hat{\theta}_{B.HT}$, using $\hat{\beta}_{B.HT}$};
     \node[punktchain, fill=green!3,join, minimum width=10cm] (baysb) {\begin{tabular}{l} Obtain the analytical form of the Bayesian MLE of\\ the intensity
function,$\hat{V}_{B.HT}$, using $\hat{\beta}_{B.HT}$ and $\hat{\theta}_{B.HT}$ \end{tabular}};
     
 \node[punktchain, ,fill=orange!8,join,minimum width=2cm] (makro) {End};

  \end{tikzpicture}
 \vspace*{3pt}
 \captionsetup{labelformat=empty}
  	\caption{\textbf{Algorithm 2.} Estimate of the intensity function using Crow data  in Table \ref{times}}. \label{alg2}
 \end{minipage}}
  \end{center}
\end{figure}

For the failure data of Crow, provided in Table \ref{times}, $\hat{\beta}_{B.HT}$ is $0.501199$ and $\hat{\theta}_{B.HT}$ is $2.07144$. Therefore, with the use of $\hat{\theta}_{B.HT}$, the Bayesian MLE of the intensity function for the data is given by:
\abovedisplayskip=20pt
\belowdisplayskip=20pt
\begin{equation}
\hat{V}_{B.HT}(t) = 0.347933 \cdot t^{-0.498801}, \: t>0.
\end{equation} 
A graphical display of $\hat{V}_{B.HT}(t)$ is given below. 

\begin{figure}[ht]
\captionsetup{width=.6\textwidth,font=footnotesize,
  justification=raggedright}
	\centerline{\includegraphics[width=10cm]{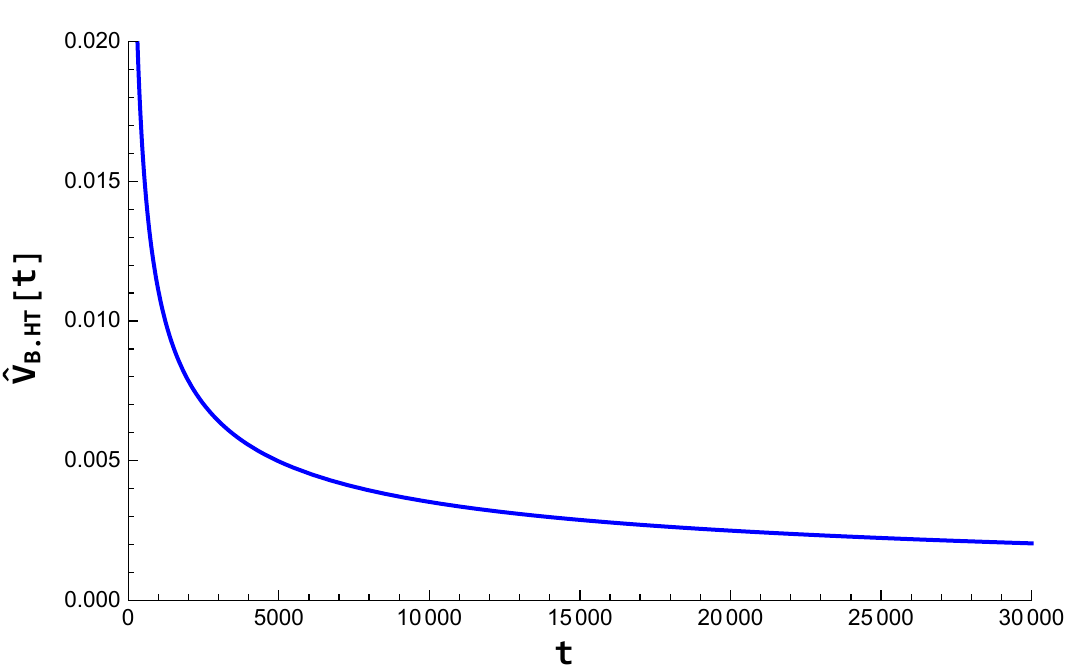}}
	\vspace*{8pt}
	\caption{Estimate of the intensity function for the real data in Table \ref{times}, using $\hat{\beta}_{B.HT}$ and $\hat{\theta}_{B.HT}$.}
    \label{intensitytime}
\end{figure}

Figure \ref{intensitytime} shows the Bayesian MLE estimate of the intensity function ($V(t;\beta,\theta)$) under the H-T loss function ($\hat{V^*}_{B.HT}$), which indicates the improvement of the software reliability over time. 

To obtain a Bayesian MLE for the reliability function under the H-T loss function, we use this Bayesian estimate for the intensity function. The analytical form for the corresponding Bayesian reliability estimate, based on the real data, is given by:
\abovedisplayskip=20pt
\belowdisplayskip=20pt
\begin{equation}
\hat{R}_{B.HT} (t_{i} | t_1,..,t_{i-1}) = \exp \left \{ -0.347933 \int ^{t_i}_{t_{i-1}}x^{-0.498801}dx \right \}, \: t_{i} > t_{i-1} > 0.
\end{equation}

Thus far, we demonstrated not only the applicability of the Bayesian analysis to the PLP, but also, using real data, the superiority of its performance and influence compared to the MLE of the parameters $\beta$ and $\theta$, respectively, assuming the Burr PDF is the prior knowledge of the key parameter $\beta$. Next section, we study the sensivity of the prior selections, in which an engineer might lack a prior knowledge of parameter $\beta$.

\subsection{Sensitivity of Prior Selection} \label{section:3.3}

In the implementation of the simulation procedure we followed Algorithm 1. Random failure times (time to failures) distributed according to the PLP are simulated for a realization of the stochastic scale parameter $\beta$, which follows a Burr type XII probability distribution. Informative parametric priors were considered, such as inverted gamma and Burr probability distributions, whereas the Jefferys prior was chosen as a non-informative prior. In addition, non-parametric priors like kernel density were applied during the sensitivity analysis study. Kernel density estimation depends on several variables, including sample size, bandwidth, and kernel function. In this study, the optimal bandwidth ($h^*$) and kernel function were chosen such that the asymptotic mean integrated squared error (AMISE) is minimized. The simplified analytical form of AMISE, Eq. (\ref{amise.f1}), is given by:
\begin{equation} \label{amise.f}
AMISE \left (\hat{f}(\beta) \right) = \frac{C(K)}{n\cdot h} + ( \frac{1}{4} \cdot h^4 \cdot k^2_2\cdot R\left(f^{(2)}(\beta)\right))
\end{equation} 
Where:
\begin{itemize}
    \item  C(K)= $\int (K(u))^{2}du$.\vspace{.2cm}
    \item n: sample size. \vspace{.2cm}
    \item h: bandwidth. \vspace{.2cm}
    \item $k_2$ = $\int_{-\infty}^{+\infty} u^2 \cdot K(u)  du$. \vspace{.2cm}
    \item $f^{(2)}(\beta)$ is the second derivative of Burr PDF. \vspace{.2cm}
    \item  $R(f^{(2)}(\beta))$= $\int (f^{(2)}(\beta))^2 d\beta$.\vspace{.2cm}
    \item  $h^* =\left[\frac{C(K)}{k^2_2 \cdot R\left(f^{(2)}(\beta)\right)}\right]^{1/5}\cdot n^{-1/5}$. 
\end{itemize} 

The minimum AMISE corresponds to the Epanechnikov kernel function ($K(u)$={\scriptsize{$\frac{3}{4}(1-u^2) I_{|u| \leq 1}$}}), \cite{412}. In addition to the Epanechnikov kernel function, the Gaussian kernel function ($K(u)$={\scriptsize {$\frac{1}{\sqrt{2\pi}}\exp\left(\frac{-u^2}{2}\right) I_{{\mathbb{I}\!R}}$}}) was also used in the calculation since it is commonly used for its analytical tractability.

\begin{figure}[!ht]
\captionsetup{width=.6\textwidth,font=footnotesize,
  justification=raggedright}
	\centerline{\includegraphics[width=10cm]{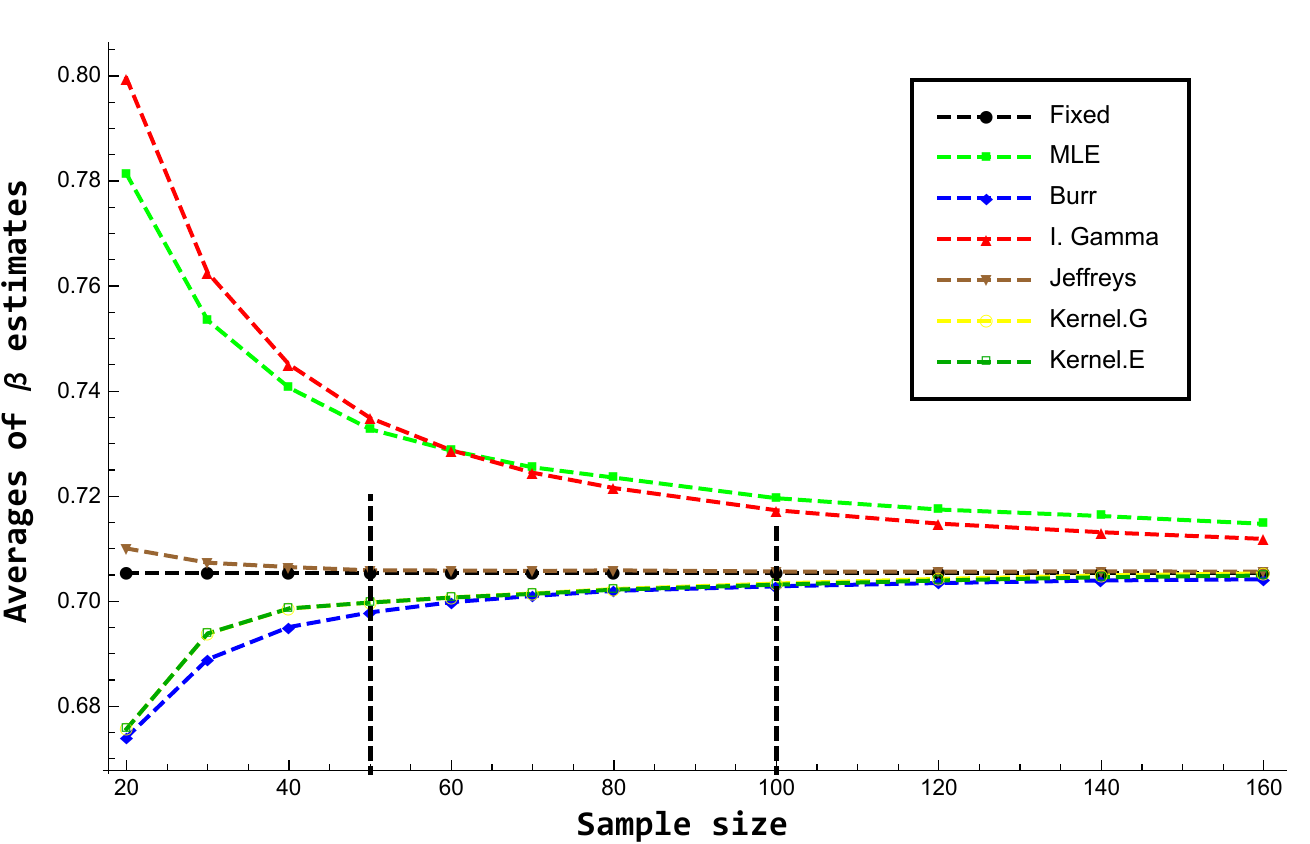}}
	\vspace*{8pt}
	\caption{Averages of $\beta$ estimates for different sample sizes.}
    \label{priorbeta1}
\end{figure}

Numerical integration techniques were used to compute the Bayesian estimates of $\beta$ under the H-T loss function, according to the equations in Section \textbf{\ref{section:2.3}}, for each of the five densities and three distinct values of $\theta$. Samples of sizes of $20$, $30$, $40$, $50$, $60$, $70$, $80$, $100$, $120$, $140$, and $160$ were generated, where the parameter $\theta$ was assumed to be the MLE of $\theta$ ($1.7441$) using Crow's data. The results, for $10,000$ repetitions, are presented by Table \ref{beta.priors} and Figure \ref{priorbeta1}.

\begin{table}[ht]
\centering
\captionsetup{width=.87\textwidth,font=footnotesize,
  justification=raggedright}
  \caption{Averages of the Bayesian estimates (using the subject prior PDFs), under H-T loss functions, and MLEs of the parameter $\beta$ over 10,000 repetitions}
{\begin{tabular}{@{}cccccccc@{}} \toprule \label{beta.priors}
\text{n} & $\beta$  & \text{$\hat{\beta}_{MLE}$} & \text{$\hat{\beta}_{B.HT}$} & \text{$\hat{\beta}_{IG.HT}$} & \text{$\hat{\beta}_{J.HT}$} & \text{$\hat{\beta}_{KG.HT}$} & \text{$\hat{\beta}_{KE.HT}$}  \\  \toprule
 20 & 0.7054 & 0.781303 & 0.674095 & 0.799535 & 0.710047 & 0.675698 & 0.675693 \\
 30 & 0.7054 & 0.753465 & 0.688951 & 0.762536 & 0.707341 & 0.693849 & 0.693881 \\
 40 & 0.7054 & 0.740665 & 0.695087 & 0.745074 & 0.70651\hphantom{0} & 0.698597 & 0.698631 \\
 50 & 0.7054 & 0.732738 & 0.697886 & 0.734888 & 0.705885 & 0.699751 & 0.69976\hphantom{0} \\
 60 & 0.7054 & 0.728669 & 0.699823 & 0.728731 & 0.705847 & 0.700727 & 0.700699 \\
 70 & 0.7054 & 0.725499 & 0.70101\hphantom{0} & 0.724471 & 0.705776 & 0.70148\hphantom{0} & 0.701405 \\
 80 & 0.7054 & 0.723539 & 0.701979 & 0.721552 & 0.705882 & 0.702323 & 0.7022\hphantom{00} \\
 100 & 0.7054 & 0.719621 & 0.70285\hphantom{0} & 0.717324 & 0.705664 & 0.70338\hphantom{0} & 0.703159 \\
 120 & 0.7054 & 0.717465 & 0.703468 & 0.71478\hphantom{0} & 0.705632 & 0.704258 & 0.703959 \\
 140 & 0.7054 & 0.716224 & 0.703954 & 0.713141 & 0.705692 & 0.704928 & 0.70457\hphantom{0} \\
 160 & 0.7054 & 0.714739 & 0.704188 & 0.711865 & 0.705629 & 0.7053\hphantom{00} & 0.704893 \\
  \bottomrule
\end{tabular}}
\end{table}

It can be observed that the Bayesian estimate of $\beta$ for the Jeffreys prior PDF under the H-T loss function converges to the true value faster than other prior PDFs; followed very closely by the Bayesian estimates using the kernel prior PDFs, which tend to converge faster than the inverted gamma prior PDF and are superior to the maximum likelihood approach. 

Figure \ref{priorbeta1} presents the graphical behavior of the MLE and Bayesian estimates represented by their average values, where the average values of the Jeffery Bayesian estimates were the closest to the true value of $\beta$ for all sample sizes. Followed by the average values of the kernel Bayesian and the Burr Bayesian estimates, which they tend to have similar behavior in estimating for sample sizes of 50 and larger. Bayesian estimates seems to have insignificant difference for sample sizes of 100 and larger except the inverted gamma Bayesian estimate which performed as poor as the MLE in estimating the key parameter $\beta$.

Table \ref{msebeeta.priorsL} shows the MSE of the Bayesian estimates of $\beta$ under the H-T loss function and the subject prior PDFs with respect to various sample sizes. \\

\begin{table}[!ht]
\centering
\captionsetup{width=1\textwidth,font=footnotesize,
  justification=raggedright}
 \resizebox{0.8\textwidth}{!}{\begin{minipage}{\textwidth}
        \caption{MSE of the Bayesian estimates, under the H-T loss functions, and MLEs of the parameter $\beta$ averaged over 1000 
repetitions for different priors.}

{\begin{tabular}{@{}ccccccc@{}} \toprule \label{msebeeta.priorsL}
\text{n}  
& \text{MSE of $\hat{\beta}_{MLE}$} & \text{MSE of $\hat{\beta}_{B.HT}$} & \text{MSE of $\hat{\beta}_{IG.HT}$} & \text{MSE of $\hat{\beta}_{J.HT}$} & \text{MSE of $\hat{\beta}_{KG.HT}$} & \text{MSE of $\hat{\beta}_{KE.HT}$}  \\  \toprule
\rowcolor{Gray}
 20 & 0.0417095\hphantom{0} & 0.00362378\hphantom{0} & 0.0115968\hphantom{00} & \textbf{0.00285493} & 0.0037072\hphantom{00} & 0.00371019\hphantom{0} \\
 30 & 0.0238951\hphantom{0} & 0.00167959\hphantom{0} & 0.00460464\hphantom{0} & 0.00139859\hphantom{0} & 0.00166486\hphantom{0} & 0.00166402\hphantom{0} \\
 40 & 0.016305\hphantom{00} & 0.00100242\hphantom{0} & 0.00242657\hphantom{0} & 0.000885863 & 0.000976117 & 0.000974269 \\
 50 & 0.0123248\hphantom{0} & 0.00067513\hphantom{0} & 0.00146266\hphantom{0} & 0.000612733 & 0.000653313 & 0.000653343 \\
 60 & 0.00997557 & 0.000502602 & 0.000998874 & 0.000467943 & 0.000476869 & 0.000478856 \\
  \rowcolor{Gray}
 70 & 0.00831476 & 0.000394076 & 0.000726943 & 0.000372566 & \textbf{0.00036205} & 0.000365206 \\
 80 & 0.00701888 & 0.000320076 & 0.000561229 & 0.000307048 & 0.000284152 & 0.000288374 \\
 100 & 0.00546558 & 0.000227799 & 0.000358957 & 0.000220557 & 0.000194561 & 0.000199899 \\
 120 & 0.00452469 & 0.000177734 & 0.000259043 & 0.000173553 & 0.000149927 & 0.000156087 \\
 140 & 0.00386396 & 0.000145074 & 0.00020085\hphantom{0} & 0.000142729 & 0.000121572 & 0.000128407 \\
  \rowcolor{Gray}
 160 & 0.00329065 & 0.000120261 & 0.00015916\hphantom{0} & 0.000118617 & \textbf{0.00010159} & 0.000108463 \\
  \bottomrule
\end{tabular}}
\end{minipage}}
\end{table}

\begin{figure}[!htbp]
\captionsetup{width=.6\textwidth,font=footnotesize,
  justification=raggedright}
	\centerline{\includegraphics[width=10cm]{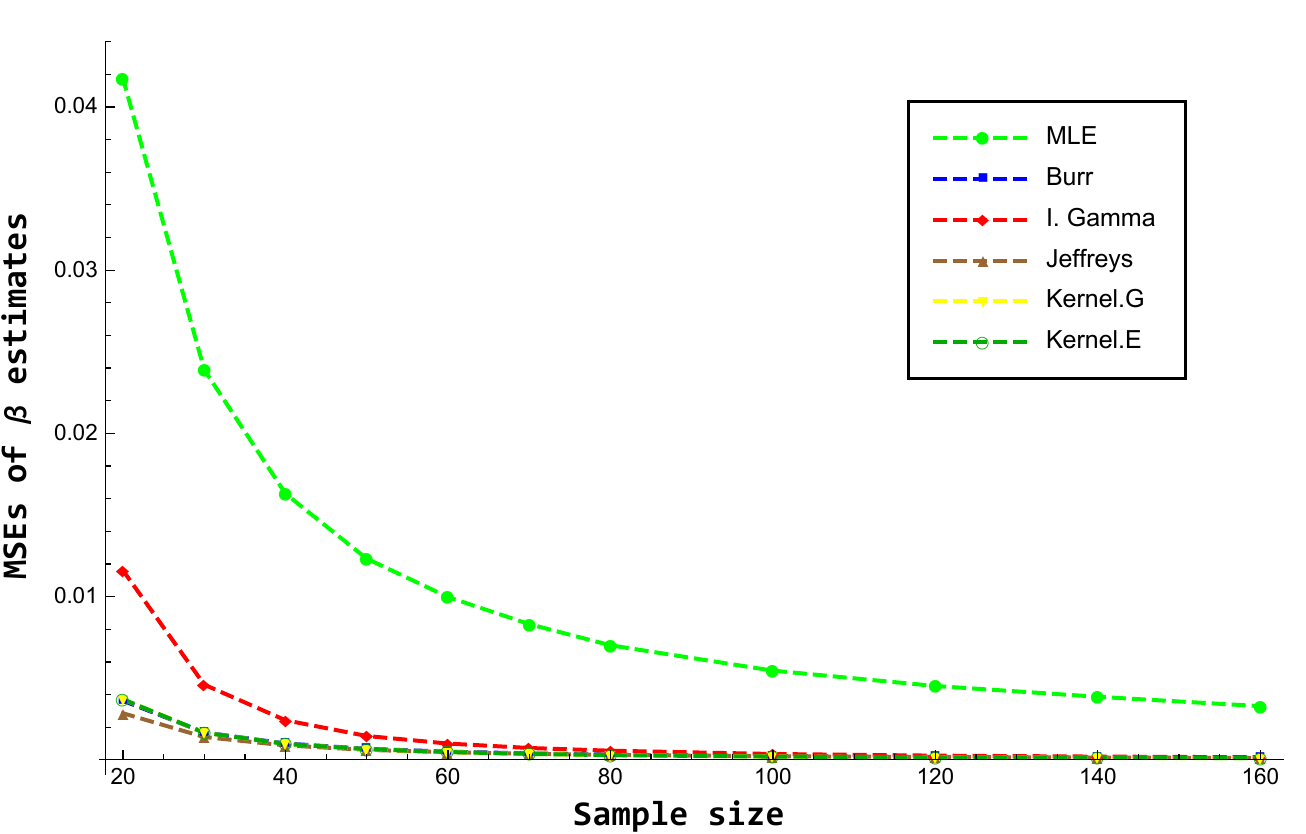}}
	\vspace*{8pt}
	\caption{MSEs of $\beta$ estimates for different sample sizes.}
    \label{msepriorbeta}
\end{figure}

\begin{figure}[!htbp]
\captionsetup{width=.6\textwidth,font=footnotesize,
  justification=raggedright}
	\centerline{\includegraphics[width=10cm]{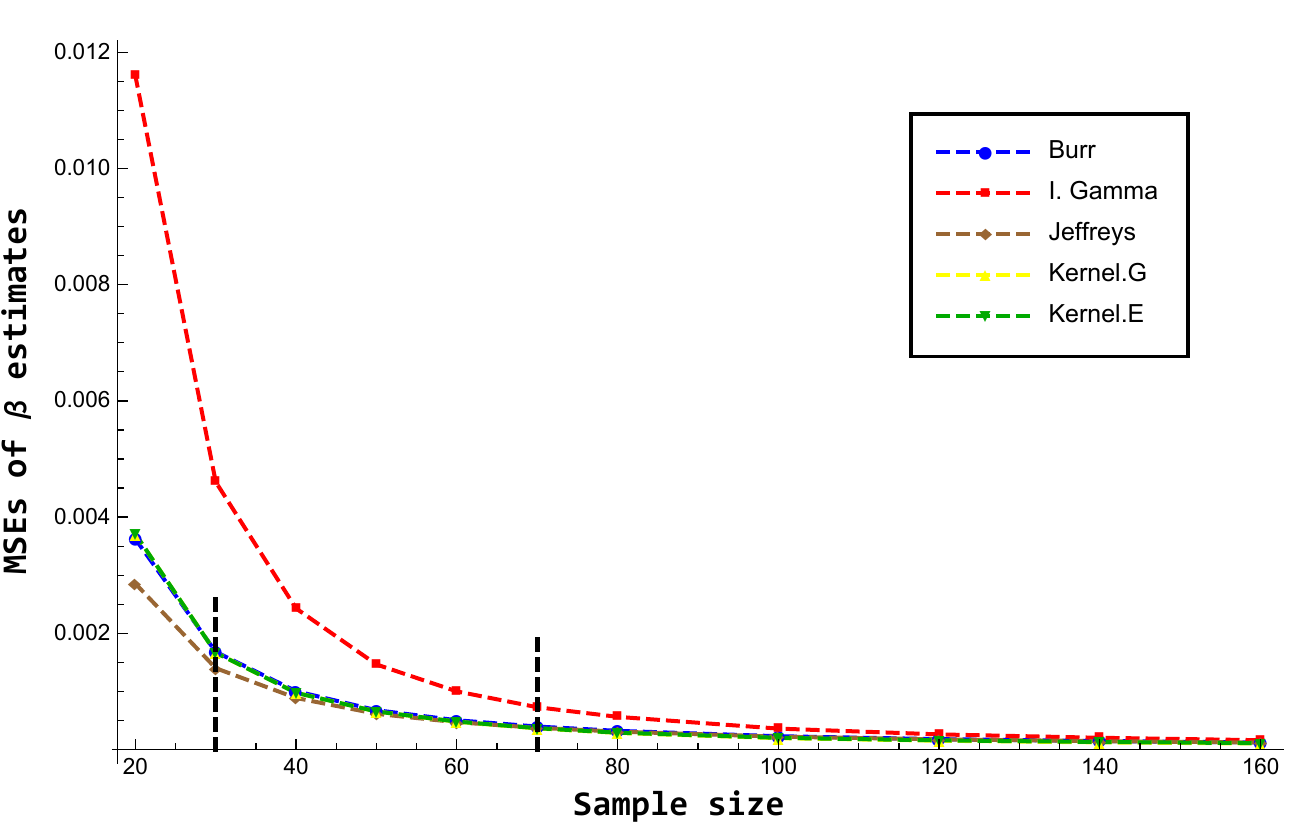}}
	\vspace*{8pt}
	\caption{MSEs of $\beta$ Bayesian estimates for different sample sizes.}
    \label{msepriorbeta2}
\end{figure}

For a sample size of 20,  the Jeffery Bayesian estimate of $\beta$ is the best estimate based on the MSE, followed by the Burr and kernel Bayesian estimates (Table \ref{msebeeta.priorsL}). The MSE of the Gaussian Kernel Bayesian esimate was the lowest for moderate to large sample sizes ($n=70,.., 160$). The inverted gamma Bayesian estimate had the lowest performance compared to other Bayesian estimates. The MSE of the MLE of the key parameter $\beta$ indicated its poor performance compared to the Bayesian estimates.  \\
Figure \ref{msepriorbeta} shows the MLE and Bayesian estimates of $\beta$. It can be observed the Bayesian estimates under the H-T loss function are superior to the MLE of the key parameter $\beta$ for all sample sizes considered in this study. Moreover, the Bayesian estimates are presented without MLE in Figure \ref{msepriorbeta2} to look closely at the performance of Bayesian estimates, under the H-T loss function and the subject prior PDFs, based on their MSEs. \\

The MSEs of the Jefferys prior, for various sample sizes, were the smallest, followed by MSEs of the kernels and Burr prior PDFs. Bayesian estimates using Gaussian and Epanechnikov kernel probability densities as priors performed similarly, whereas the Bayesian estimate using the inverted gamma PDF had the lowest performance compared to other Bayesian estimates. It also shows that for a sample size of $n=30$ and larger, the Bayesian estimates using Burr, Jeffery, and kernel PDFs are similar in their convergence to the true value. For a sample size of $n=70$ and larger, all Bayesian estimates tend to converge to the true value of the key parameter $\beta$.
The Bayesian estimates of $\beta$ under Jefferys and inverted gamma prior PDFs tend to overestimate $\beta$, whereas Burr and kernel prior PDFs tend to underestimate $\beta$.

For each sample of size 40 based on Monte Carlo simulation, the Bayesian estimates and MLEs of 
$\beta$ were calculated when $\theta \in \{0.5, 1.7441, 4\}$. The comparison is based on the MSE averaged over the $2,000$ simulated samples. The results are given by Table \ref{Mse.Beta.theta.priorsL}. 

\begin{table}[!ht]
\centering
\captionsetup{width=1\textwidth,font=footnotesize,
  justification=raggedright}
\resizebox{0.75\textwidth}{!}{\begin{minipage}{\textwidth}
\caption{MSE of $\beta$ Bayesian estimates with Burr, Jeffreys, inverted gamma, and kernel PDFs as priors under the
H-T loss function. MSE of MLE estimate of the parameter $\beta$ in an NHPP with
samples of n = 40 and different values of the parameter $\theta$.}
{\begin{tabular}{@{}ccccccc@{}} \toprule \label{Mse.Beta.theta.priorsL}
 $\theta$ & \text{MSE of $\hat{\beta}_{MLE}$} & \text{MSE of $\hat{\beta}_{B.HT}$} & \text{MSE of $\hat{\beta}_{IG.HT}$} & \text{MSE of $\hat{\beta}_{J.HT}$} & \text{MSE of $\hat{\beta}_{KG.HT}$} & \text{MSE of $\hat{\beta}_{KE.HT}$}   \\  \toprule
 0.5 & 0.0106298\hphantom{0} & \textbf{0.000520244} & 0.00229073 & 0.000577579 & 0.000535304\hphantom{0} & 0.000534979 \\
 1.7441 & 0.00996687 & 0.000524716\hphantom{0} & 0.00225675 & 0.000585273 & \textbf{0.000516497} & 0.000516599 \\
 4 & 0.0112937\hphantom{0} & 0.000558022\hphantom{0} & 0.00246984 & 0.000632173 & \textbf{0.000556695} & 0.000556804 \\
  \bottomrule
\end{tabular}}
\end{minipage}}
\end{table}

It can be observed that all Bayesian estimates are superior to MLE ($\hat{\beta}$) in estimating $\beta$, with sample size n = 40, while maintaining a consistent behavior for the different values of $\theta$ (Table \ref{Mse.Beta.theta.priorsL}). For small value of $\theta$, Jeffrey Bayesian estimate had the lowest MSE value, whereas the MSE of the Gaussian kernel Bayesian estimate was the lowest for moderate and large values of $\theta$. \\ 
For the case in which we misleadingly assumed the true probability distribution of the key parameter $\beta$, we found that the Jeffreys and kernel Bayesian estimates of $\beta$ had the best performance compared to the inverted gamma Bayesian estimate of $\beta$, indicating that the Bayesian estimate of $\beta$ is sensitive to the choice of its prior PDF.

\begin{figure}[!ht]
\captionsetup{width=.6\textwidth,font=footnotesize,
  justification=raggedright}
	\centerline{\includegraphics[width=10cm]{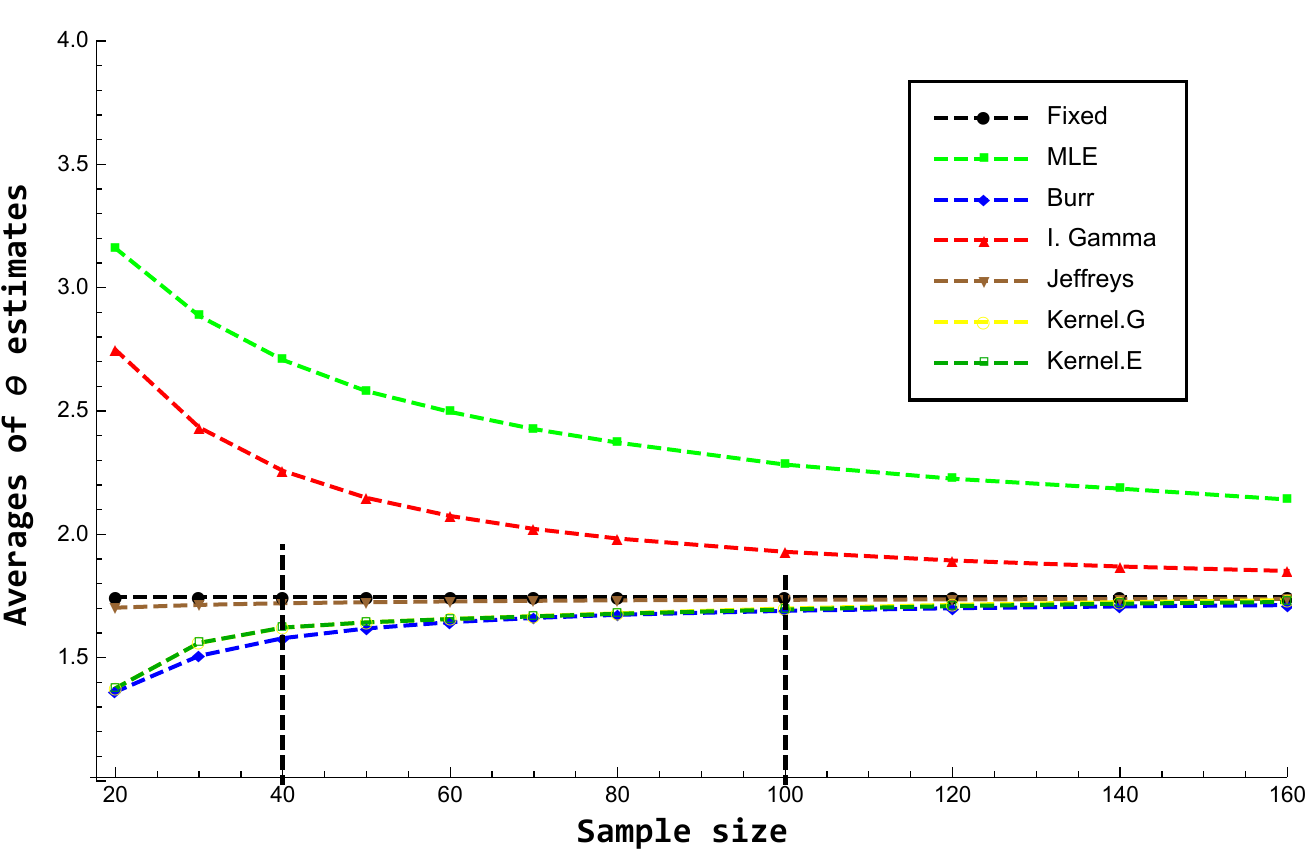}}
	\vspace*{8pt}
	\caption{Averages of MLEs of $\theta$ using the Bayesian estimates of the parameter $\beta$ with respect to different priors.}
    \label{priortheta1}
\end{figure}

As expected, the adjusted MLE of $\theta$ produced a better estimate under the mentioned
Bayesian influence with respect to the H-T loss function. The average values of the MLE and Bayesian estimates of $\theta$ using the MLE and Bayesian estimates of $\beta$ for various sample sizes, respectively, are presented by Figure \ref{priortheta1}. where the average values of the Jeffery Bayesian estimates were the closest to the true value of $\theta$ for all sample sizes. Followed by the average values of the kernel Bayesian and the Burr Bayesian estimates, which tend to have similar behavior in estimating for sample sizes of 40 and larger. When considering sample sizes of 100 and larger, the Bayesian estimates seems to have insignificant difference except for the inverted gamma Bayesian estimate which which still performs better than the MLE in estimating the parameter $\theta$.

The MSE of $\theta$ estimates using MLE and Bayesian estimates of $\beta$, namely $\hat{\theta}_{MLE}$, $\hat{\theta}_{B.HT}$, $\hat{\theta}_{IG.HT}$, $\hat{\theta}_{J.HT}$, $\hat{\theta}_{KG.HT}$, and $\hat{\theta}_{KE.HT}$, are shown in Table \ref{msetheta.priorsL} with various sample sizes. 
For a small sample, moderate, and large sample sizes, $n=20,70,160$ respectively, the Jeffrey Bayesian estimate of $\theta$ ($\hat{\beta}_{J.HT}$) performed better than the other estimates, followed by the kernel Bayesian and Burr Bayesian estimates. 

\begin{table}[!htbp]
\centering
\captionsetup{width=1\textwidth,font=footnotesize,
  justification=raggedright}
\resizebox{0.8\textwidth}{!}{\begin{minipage}{\textwidth}
\caption{MSE of the Bayesian estimates, under the H-T loss functions, and MLEs of the parameter $\theta$ averaged over 1000
repetitions for different priors.}
{\begin{tabular}{@{}ccccccc@{}} \toprule \label{msetheta.priorsL}
\text{n}  
& \text{MSE of $\hat{\theta}_{MLE}$} & \text{MSE of $\hat{\theta}_{B.HT}$} & \text{MSE of $\hat{\theta}_{IG.HT}$} & \text{MSE of $\hat{\theta}_{J.HT}$} & \text{MSE of $\hat{\theta}_{KG.HT}$} & \text{MSE of $\hat{\theta}_{KE.HT}$}  \\  \toprule
  \rowcolor{Gray}
 20  & 11.5735 & 0.175991\hphantom{00} & 1.0789\hphantom{000} & \textbf{0.0310651} & 0.174974\hphantom{00} & 0.175081\hphantom{00} \\
 30 & 8.6721\hphantom{0} & 0.0726953\hphantom{0} & 0.50118\hphantom{00} & 0.0166961\hphantom{0} & 0.0538952\hphantom{0} & 0.0537907\hphantom{0} \\
 40 & 6.67926 & 0.0378416\hphantom{0} & 0.27733\hphantom{00} & 0.0107884\hphantom{0} & 0.0264298\hphantom{0} & 0.026306\hphantom{00} \\
 50 & 5.45282 & 0.0233418\hphantom{0} & 0.171955\hphantom{0} & 0.00777231 & 0.0182648\hphantom{0} & 0.018213\hphantom{00} \\
 60  & 4.55577 & 0.015811\hphantom{00} & 0.115456\hphantom{0} & 0.00593216 & 0.0139622\hphantom{0} & 0.0139218\hphantom{0} \\
 \rowcolor{Gray}
 70 & 3.92543 & 0.0114843\hphantom{0} & 0.0820851 & \textbf{0.0047144} & 0.0110305\hphantom{0} & 0.0109659\hphantom{0} \\
 80 & 3.38362 & 0.00867134 & 0.0610179 & 0.00382098 & 0.00880575 & 0.00869148 \\
 100  & 2.73005 & 0.00566115 & 0.0369537 & 0.0027899\hphantom{0} & 0.00591856 & 0.00572456 \\
 120  & 2.33339 & 0.00402454 & 0.0245551 & 0.00215943 & 0.004114\hphantom{00} & 0.00387668 \\
 140  & 2.03741 & 0.00302931 & 0.0174259 & 0.00174116 & 0.003041\hphantom{00} & 0.00277349 \\
  \rowcolor{Gray}
 160 & 1.75235 & 0.00238185 & 0.0129141 & \textbf{0.0014346} & 0.0023649\hphantom{0} & 0.00206666 \\
  \bottomrule
\end{tabular}}
\end{minipage}}
\end{table}

\begin{figure}[ht]
\captionsetup{width=.6\textwidth,font=footnotesize,
  justification=raggedright}
	\centerline{\includegraphics[width=10cm]{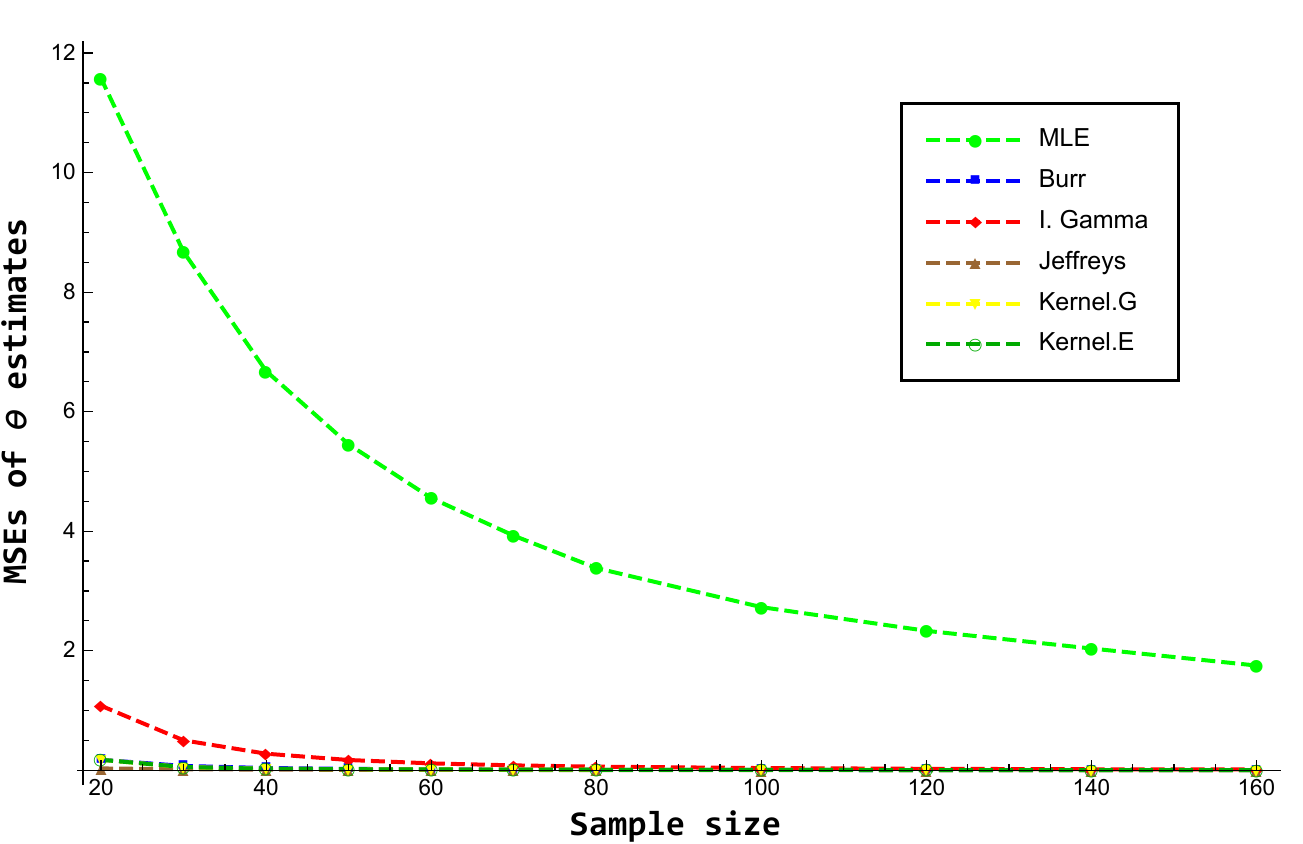}}
	\vspace*{8pt}
	\caption{MSE of the MLEs of $\theta$ using MLE and Bayesian estimates of $\beta$ with respect to different prior $\beta$}
    \label{msethetaL1}
\end{figure}
\noindent
While the inverted gamma Bayesian estimate of $\theta$ outperformed that of the MLE, it
did not perform as well as the other Bayesian estimates based on their MSE values
across all sample size. This indicates the MLE of $\theta$ had the poorest performance compared to the Bayesian estimates. 

\begin{figure}[!ht]
\captionsetup{width=.6\textwidth,font=footnotesize,
  justification=raggedright}
	\centerline{\includegraphics[width=10cm]{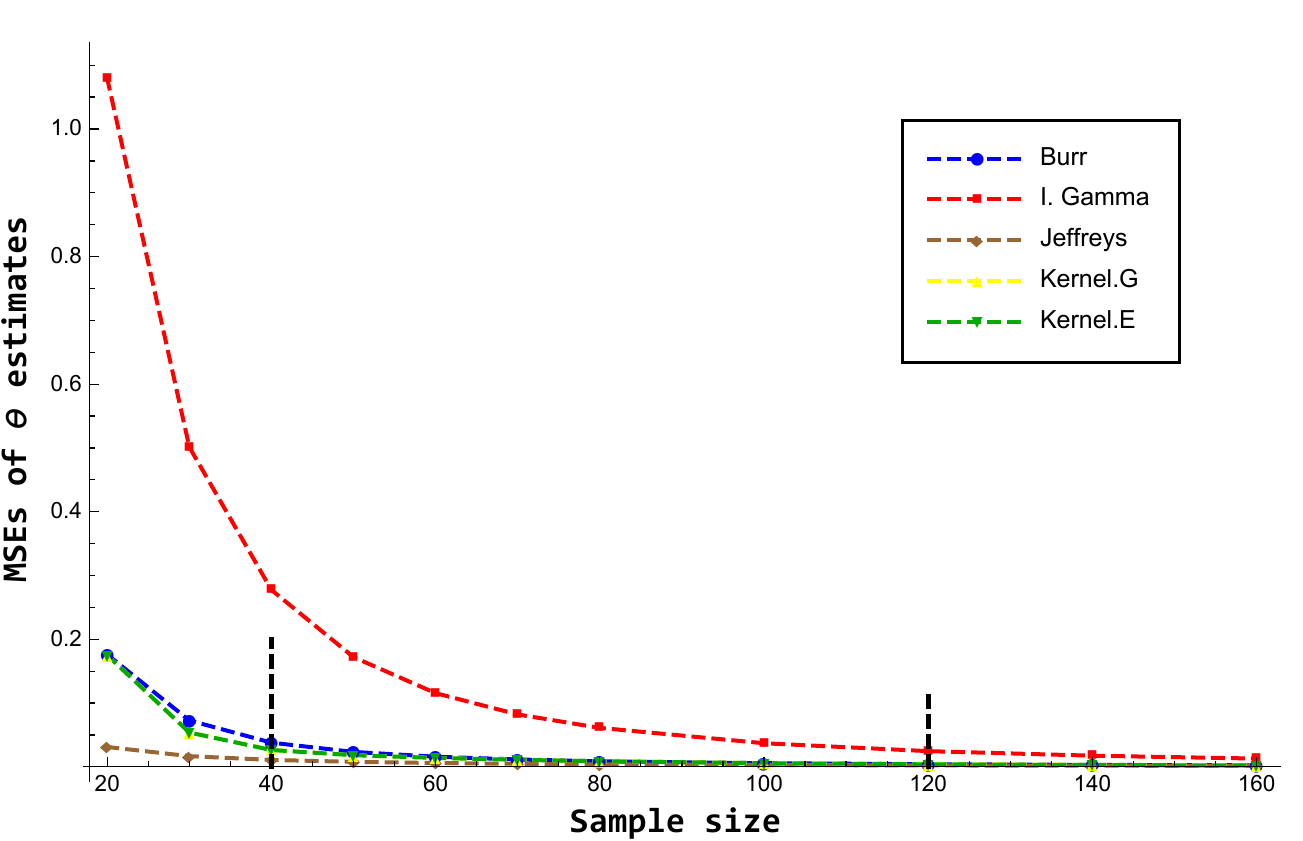}}
	\vspace*{8pt}
	\caption{MSE of the MLEs of $\theta$ using Bayesian estimates of $\beta$ with respect to different prior $\beta$.}
    \label{msethetaL3}
\end{figure}

The MSEs of $\theta$ estimates using the MLE and Bayesian estimates of the parameter $\beta$ with respect to different priors is displayed in Figure \ref{msethetaL1}, from which it can be seen that the MLE of $\theta$ was extrmely weak estimator since it has the largest MSEs across the sample sizes. The adjusted $\theta$ estimates were were displayed without the MLE estimates by Figure \ref{msethetaL3}. 

It can be noted that the $\theta$ estimate using the inverted gamma Bayesian estimate of $\beta$ had the lowest performance compared to other estimates that used Bayesian estimates of $\beta$. In addition, above the sample size $n=40$, the estimate of $\theta$ using Burr kernels Bayesian estimates of $\beta$ performed similarly in estimating the true value of $\theta$. All $\theta$ estimates using Bayesian estimates of $\beta$ tend to converge to the true value in similar trajectories, whereas the $\theta$ estimate using the Jeffrey Bayesian estimate of $\beta$ converges slightly faster. 

Therefore, the Bayesian analysis to the PLP under the H-T loss function is sensitive to the
prior selection. Based on this finding, an engineer, for example, is recommended to use the Jeffreys or kernel PDFs if he/she lacks any prior knowledge of $\beta$.

\section{Conclusion}

In the present study, we developed the analytical Bayesian form of the key parameter $\beta$, under the H-T loss function, in the intensity function, where the underlying failure distribution is the PLP that is used for software reliability assessment, among others. The reliability function of the subject model is written analytically as a function of the intensity function.

The behavior of  $\beta$ is characterized by the Burr type XII probability distribution. Real data and numerical simulation were used to illustrate the efficiency improvement in the estimation of the intensity function of PLP under the H-T loss function ($\hat{V}_{B.HT}(t)$). 
Based on the $100,000$ samples of software failure times, using Monte Carlo simulations and sample size of 40, we found that the Bayesian estimate of $\beta$ under the H-T loss function ($\hat{\beta}_{B.HT}$) performed better than the MLE of $\beta$ with respect to three different values of $\theta$ ( 0.5, 1.7441, 4 ). Even for different sample sizes ($20$, $30$, $40$, $50$, $60$, $70$, $80$, $100$, $120$, $140$, and $160$), similar results were achieved using $\beta$=0.7054, $\theta$=1.7441, and averaged over $10,000$ samples of software failure times. 

Because the MLE of the second parameter ($\theta$) in the intensity function depends on the estimate of $\beta$, the adjusted estimate of $\theta$  $\hat{\beta}_{B.HT}$, as expected, performed better than MLE of $\theta$. Moreover, by comparing the relative efficiency metric, mainly using MLEs for both $\beta$ and $\theta$ ($\hat{V}_{MLE}(t)$), using Bayesian estimates of $\beta$ under the H-T loss function, and Bayesian MLE of $\theta$ ($\hat{V}_{B.HT}(t)$), we found that $\hat{V}_{B.HT}(t)$ is more efficient in estimating the intensity function $V(t;\beta,\theta)$.

In section \textbf{\ref{section:3.3}}, we answered the second research question: Is the Bayesian estimate of the key parameter, $\beta$, using the H-T loss function in the PLP, sensitive to the selection of the prior, whether parametric or non-parametric? The parametric priors were Burr, Jefferys, and inverted gamma probability distributions.  The non-parametric priors were the Gaussian and Epanechnikov kernel densities. The priors' parameters were estimated using Crow failure times. Additionally, the optimal bandwidth and kernel functions were selected to minimize the asymptotic mean integrated squared error. 

Using the proposed algorithm, the Bayesian estimate of $\beta$ under the H-T loss function and Burr PDF as a prior, $\hat{\beta}_{B.HT}$, performed slightly better than the other Bayesian estimates using different prior PDFs for small value of $\theta$. The Bayesian estimate of $\beta$ under H-T loss function and Gaussian kernel PDF as a prior, $\hat{\beta}_{KG.HT}$, had the smallest MSE comparing to other prior PDFs for moderate and large values of $\theta$. MLE of $\beta$ continued the poor performance comparing to Bayesian estimates using the subject prior PDFs. The Bayesian estimate of $\beta$ under H-T loss function and inverted gamma PDF as a prior performed the lowest among other Bayesian estimates using Burr, Jeffrey, Gaussian, and Epanechnikov PDFs as priors whereas the latter have slightly different MSEs values. Even for different sample sizes, the MLE of $\beta$ has the poorest performance comparing to Bayesian estimates using different prior PDFs. For small to moderate sample sizes ($n=20, 30, 40, 60$), the Bayesian estimate of $\beta$ under H-T loss function and Jeffrey PDF as a prior, $\hat{\beta}_{J.HT}$, has the smallest MSE value, followed closely by the Bayesian estimates using the Gaussian and Epanechnikov kernels, and Burr PDFs as priors. For moderate to large sample sizes ($n=70, 80,.., 160$), the Bayesian estimate of $\beta$ under H-T loss function and Gaussian kernel PDF as a prior, $\hat{\beta}_{KG.HT}$, has the smallest MSE value, followed by the Byesian estimates using the Epanechnikov kernel, Jeffrey, and Burr PDFs as priors. 

The Bayesian estimates under H-T loss function and the parametric and non-parametric priors were used the compute the adjusted estimate of $\theta$, where the adjusted $\theta$ estimate using the Jeffery Bayesian estimate of $\beta$ performed the superior estimate comparing to MLE and other Bayesian estimates for various sample sizes. Followed by Bayesian estimates using Epanechnikov kernel, Burr, and Gaussian kernel PDFs as priors for moderate to large sample sizes whereas using the Gaussian kernel Bayesian of $\beta$ to compute the adjusted estimate of $\theta$ performed slightly better comparing to the usage of Burr Bayesian estimate of $\beta$.

Thus, based on this aspect of our analysis, we can conclude that the Bayesian analysis approach under Higgins-Tsokos loss function is superior to the maximum likelihood approach in estimating the reliability function of the Power Law Process. Therefore, the results of this study have the potential to contribute not only to the reliability analysis field but also to other fields that employ the Power Law Process.


\bibliographystyle{plain}

\clearpage


\end{document}